\documentclass[preprint,12pt]{elsarticle}

\usepackage{amssymb}
\usepackage{amsmath}

\usepackage[most]{tcolorbox}

\usepackage[hyphens,spaces,obeyspaces]{url}
\usepackage[colorlinks,allcolors=blue]{hyperref}

\usepackage{cleveref}
\usepackage{physics} 
\usepackage{lmodern}

\usepackage{tikz} 
\usetikzlibrary{arrows.meta} 
\usetikzlibrary{decorations.markings}
\usetikzlibrary{backgrounds}
\usetikzlibrary{positioning,chains,fit,shapes,calc}
\usetikzlibrary{angles,patterns}
\usetikzlibrary{matrix,quotes}
\usetikzlibrary{decorations.pathmorphing, calc}
\usetikzlibrary{mindmap}
\usetikzlibrary{shapes.geometric}
\usepackage{circuitikz}

\definecolor{aliceblue}{rgb}{0.94, 0.97, 1.0}

\newcommand{\enc}{\text{Enc}}
\newcommand{\dec}{\text{Dec}}

\newtheorem{definition}{Definition}%

\begin{document}

\begin{frontmatter}



\title{Privacy Technologies for Financial Intelligence}


\author[inst1]{Yang Li\corref{cor1}}
\ead{kelvin.li@deakin.edu.au}

\author[inst2]{Thilina Ranbaduge}
\ead{thilina.ranbaduge@data61.csiro.au}

\author[inst3]{Kee Siong Ng}
\ead{keesiong.ng@anu.edu.au}

\cortext[cor1]{Corresponding author: Yang Li}

\affiliation[inst1]{organization={School of Information Technology, Deakin University},
                    state={VIC},
                    country={Australia}}

\affiliation[inst2]{organization={Data61, CSIRO},
                    state={ACT},
                    country={Australia}}

\affiliation[inst3]{organization={School of Computing, Australian National University},
                    state={ACT},
                    country={Australia}}

\begin{abstract}
Financial crimes like money laundering and terrorism financing can have significant impacts on society, including loss of trust in the integrity of the financial system, misuse and mismanagement of public funds, increase in societal problems like drug trafficking and illicit gambling, and loss of innocent lives due to terrorism activities. 
Effective detection of complex financial crimes remains a formidable challenge for regulators and financial institutions because the critical data needed to establish patterns and criminality are often dispersed across multiple organisations and cannot be linked due to privacy constraints around large-scale data matching.
Recent advances in privacy and confidential computing technologies, which enable private and secure data analysis across organisations, offer a promising opportunity for regulators and the financial industry to come together to enhance their collaborative risk detection while maintaining privacy standards. 
This paper, through a survey of the financial intelligence ecosystem, seeks to identify opportunities for the utilisation of privacy technologies to improve the state-of-the-art in financial-crime detection.
\end{abstract}




\end{frontmatter}



\section{Introduction}\label{sec1}

A robust and trustworthy banking and financial system is essential for the stability and proper functioning of any mature economy. However, in recent years, the reputation and integrity of several major financial institutions have been seriously undermined as organised criminal networks have exploited loopholes and systemic vulnerabilities to launder vast sums of money, some of which have even been linked to terrorist financing \cite{lynch22, leuprecht23}. 
Beyond the reputational harm inflicted on financial institutions, sophisticated financial crimes like money laundering and terrorism financing have profound societal consequences, including:
\begin{itemize}\itemsep1mm\parskip0mm
 \item mismanagement and misuse of public funds;
 \item financial losses for institutions due to increased compliance costs and regulatory penalties;
 \item escalation of societal issues such as drug trafficking, human smuggling, and illegal gambling, which carry significant economic burdens;
 \item the tragic loss of innocent lives in the case of terrorist activities.
\end{itemize} 
Globally important cases like Binance's \$4 billion fine\footnote{\href{https://www.justice.gov/archives/opa/pr/binance-and-ceo-plead-guilty-federal-charges-4b-resolution}{Binance's \$4 billion fine}} for failing to maintain an effective anti-money-laundering (AML) program and other violations against the US Bank Secrecy Act, UBS's \$4.2 billion fine by Swiss, US and French regulators for facilitating tax evasion and inadequate AML controls, and Goldman Sachs's \$2.9 billion fine\footnote{\href{https://www.justice.gov/usao-edny/pr/goldman-sachs-resolves-foreign-bribery-case-and-agrees-pay-over-29-billion}{Goldman Sachs's \$2.9 billion fine}} for the 1MDB money-laundering and bribery violations, have certainly brought these issues into the public consciousness and discourse.
In Australia, the government's successful court proceedings against the Commonwealth Bank of Australia, Westpac Banking Corporation, and Crown Casinos for serious breaches of the AML/CTF Act, and their ongoing legal actions against the operators of the Star Casinos and 
online betting sites like Ladbrokes and Neds highlight the continued challenges faced by regulators and financial intelligence agencies in combating financial crimes.\footnote{\href{https://www.austrac.gov.au/lists-enforcement-actions-taken}{Lists of enforcement actions taken}}

In response to these challenges, AUSTRAC, Australia's Financial Intelligence Agency, established the Fintel Alliance\footnote{\href{https://www.austrac.gov.au/partners/fintel-alliance}{Fintel Alliance}} in 2016 to foster collaboration between government agencies, financial institutions, and law enforcement. 
A number of similar initiatives have also emerged globally, including the UK's Joint Money Laundering Intelligence Task Force\footnote{\href{https://www.nationalcrimeagency.gov.uk/what-we-do/national-economic-crime-centre}{UK's Joint Money Laundering Intelligence Task Force}} and the U.S. Financial Crime Enforcement Network Exchange\footnote{\href{https://www.fincen.gov/resources/financial-crime-enforcement-network-exchange}{U.S. Financial Crime Enforcement Network Exchange}}. These partnerships have proven effective in addressing known threats, but they still face formidable challenges in detecting complex financial crimes because the critical data needed to establish patterns and criminality are often dispersed across multiple organisations and cannot be linked due to privacy constraints around large-scale data matching \cite{leuprecht18}.
The problem is further compounded by the growing complexity of emerging value-transfer mechanisms such as cryptocurrencies and real-time payment systems.

Recent innovations in Privacy-Preserving Data Matching and Machine Learning, which enable private and secure data analysis across organisations, offer a promising opportunity for regulators and the financial industry to come together to enhance their collaborative risk detection while maintaining privacy standards. 
This paper, through a survey of the financial intelligence ecosystem, seeks to identify opportunities for the utilisation of privacy technologies to improve the state-of-the-art in financial-crime detection. 
Our paper complements the material covered in \cite{ffis2021}, which explored global case studies of public and private sectors employing 
privacy technologies to foster collaborative financial intelligence.  

\section{Financial Intelligence}\label{sec:fintel}
Financial intelligence is a highly specialised field that is not well
understood by many. This section provides an introduction to the financial 
intelligence ecosystem, including a description of the 
current state, some existing limitations and opportunities.

\subsection{The Financial Intelligence Ecosystem}
\begin{figure}
    \centering
    \includegraphics[width=1.0\textwidth]{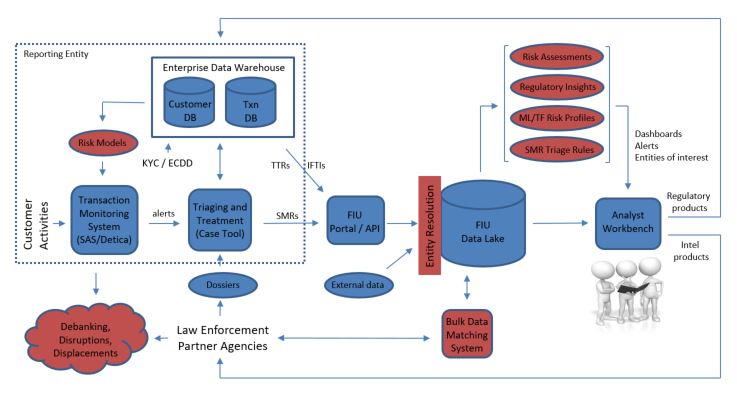}
    \caption{The figure shows a simplified schematic view of a financial intelligence ecosystem. 
    }
    \label{fig:ecosystem}
\end{figure}

Financial Intelligence Units (FIUs) around the world are governed by 
anti-money laundering and counter-terrorism financing (AML/CTF) regimes
that are quite similar in nature \cite{madinger2011money, gilmore04}.
Reporting Entities (REs) are companies that provide designated financial services that are regulated under AML/CTF laws and typically include banks, money remitters, digital currency exchanges, casinos, and increasingly law firms and real-estate companies.
(In a country the size of Australia, there are about 10,000 REs in the financial system.) 
These REs are required by law to maintain know-your-customer (KYC) and anti-money laundering programs to make sure they can verify their customers' identities and prevent their designated services from being used for illicit purposes.
Certain types of customer transaction data, including threshold transaction reports (TTRs) for cash transactions exceeding \$10k and all international fund transfer instructions (IFTIs) regardless of amount, are automatically sent to the relevant FIUs on a regular basis.
In addition, REs can also submit suspicious matter reports (SMRs) or suspicious activity reports (SARs) to FIUs if they detect suspicious customer activities through their internal controls.  
Adding to all these data from REs, FIUs may also get reports about specific individuals or entities of interest from law-enforcement agencies and partner FIUs in other countries. 
Using all these collected data, analysts in FIUs produce 
intelligence products that either report on specific threats that
can be actioned by law enforcement agents, or describe macro threats and 
mitigation strategies that financial institutions should adopt in 
their internal controls. Figure \ref{fig:ecosystem} shows the 
information flow from reporting entities to FIUs and, from there, to 
law-enforcement agencies and the limited feedback loops in the system.

As we can see, there are risk-monitoring and risk-reduction mechanisms built into many different parts of the financial intelligence ecosystem, both within reporting entities in a distributed way and more centrally within the FIU.
Many of these organisations also have large-scale automated transaction-monitoring systems that are augmented by manual analyst investigations and formal governance oversight. 
The parts of Figure~\ref{fig:ecosystem} coloured in brown represent areas of the financial intelligence ecosystem where artificial intelligence and machine learning can have a significant impact. This is primarily because the sheer quantity of data that must be processed in these locations is immense, and the connections and relationships that need to be distilled and understood, including causal and deliberately hidden linkages, are often complex and intricate.

\begin{figure}
\begin{center}
    \includegraphics[width=0.75\textwidth]{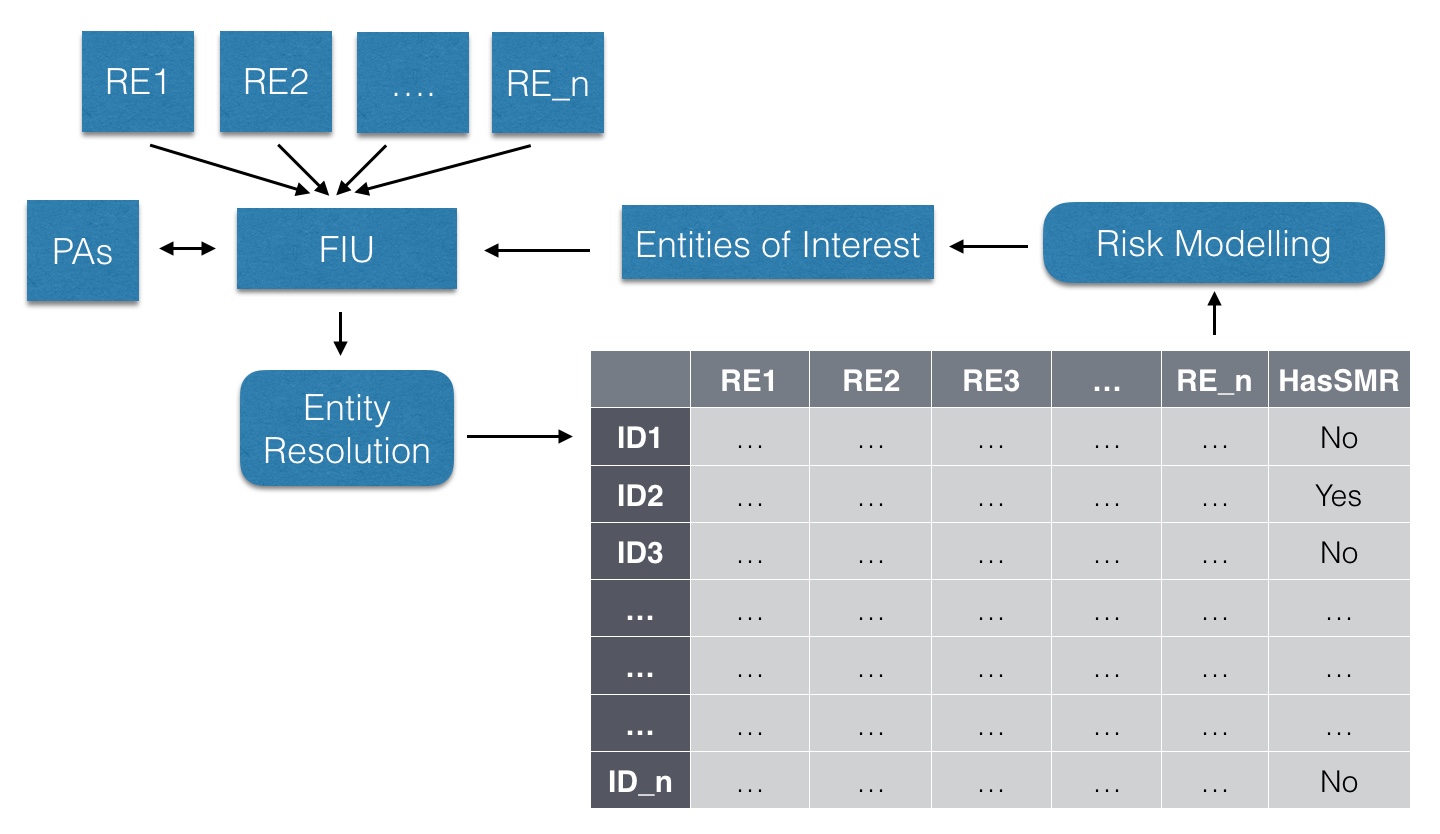}
    \caption{The entity-resolved data and risk indicators available to a typical financial intelligence agency.}
    \label{fig:fc-current}
\end{center}
\end{figure}

Broadly speaking, all data received by an FIU are typically first processed through an entity-resolution step to identify and consolidate records that are highly probable to relate to the same underlying entity \cite{zhang18}. 
The aim is to produce a dataset like that shown in Figure~\ref{fig:fc-current}, where each row represents an entity (either persons, corporations, or clusters of related individuals/companies) and the columns illustrate their activities within the financial ecosystem as observed through the lens of the financial institutions whose services they engage with.
A small set of such entities would also have explicit risk indicators flagged against them based on, for example, suspicious matter reports, sanctions list and other criminal intelligence data. 
Given a dataset like Figure~\ref{fig:fc-current}, an FIU equipped with appropriate technical expertise and advanced data-analytics platforms can employ profiling and statistical machine learning methodologies, both supervised and unsupervised, at varying levels of granularity from individual transactions to single entities and networks of entities, to try to uncover potential risks in the financial system and take appropriate action. 
Examples of such work include \cite{savageWCZY16,savageWZCY17,savage-tpds-17}.

A crucial aspect to understand about financial intelligence is that financial crimes cannot be definitively identified solely from financial transaction data. Here, at a high level, are the types of anomalies we can infer from financial transaction data \cite{madinger2011money, reuter04}:
\begin{enumerate}\itemsep1mm\parskip0mm
\item the origin of funds cannot be explained or is obscure;
\item there are rapid transfers of money through various channels, potentially indicating an attempt to obscure the money trail;
\item the financial transaction patterns are inconsistent with a customer's anticipated personal or business profile.
\end{enumerate}
It is important to note that these activities, in isolation, are not inherently illegal. Criminality can only be established and addressed when these activities are linked to a predicate crime, such as tax evasion, welfare-payment fraud, drug trafficking, or similar crimes.
To ensure the entire system functions effectively, law enforcement presence is essential throughout the financial intelligence ecosystem to provide leads and supply appropriate risk contexts. 
Specifically, pure discovery work within financial transaction data presents considerable challenges, but commercial databases of entities of interest, such as LSEG World-Check, and external datasets like company registration information can prove valuable in adding context to financial data.

That is the current state.
The overall system is not perfect by any means. Nevertheless the regime is
reasonably effective in deterring, detecting, and disrupting financial crimes, 
as demonstrated by the many successful prosecutions listed in Section~\ref{sec1}.
The main limitation of the current state is in the links that are there in the physical financial system but remain missing in the financial intelligence ecosystem. 
To begin with, a lot of financial transaction data, including domestic 
interbank transfers and crypto currency transactions, are not reported by default to an FIU and that is a major blind spot for analysts. 
Compounding the issue, the many thousands of reporting entities in a mature economy like Australia 
are heavily handicapped in their ability to share intelligence and data with each other because of 
legal impediments and risks around accidental tip-offs. 
One can also argue that there is simply not enough feedback loops built into the system, including
that between reporting entities and an FIU, and that
between an FIU and its partner law-enforcement agencies and foreign FIUs. 

In summary, the key limitations in the current state of financial intelligence can be traced
to the following issues:
\begin{itemize}\itemsep1mm\parskip0mm
    \item Sophisticated and complex financial crimes typically involve transactions that 
    span multiple financial institutions in multiple geographies, going 
    through multiple payment channels, some of which remain unregulated.
    \item Criminality cannot be established through analysis of financial transaction data alone; actual links to predicate crimes are necessary.
   \item FIUs, reporting entities, and law-enforcement agencies are all mostly working within data silos and it is hard for any one organisation to get a complete view of financial and criminal activities across the entire system.
\end{itemize}

\subsection{A Way Forward using Privacy Technologies}

\begin{figure}
    \centering
    \includegraphics[width=0.75\textwidth]{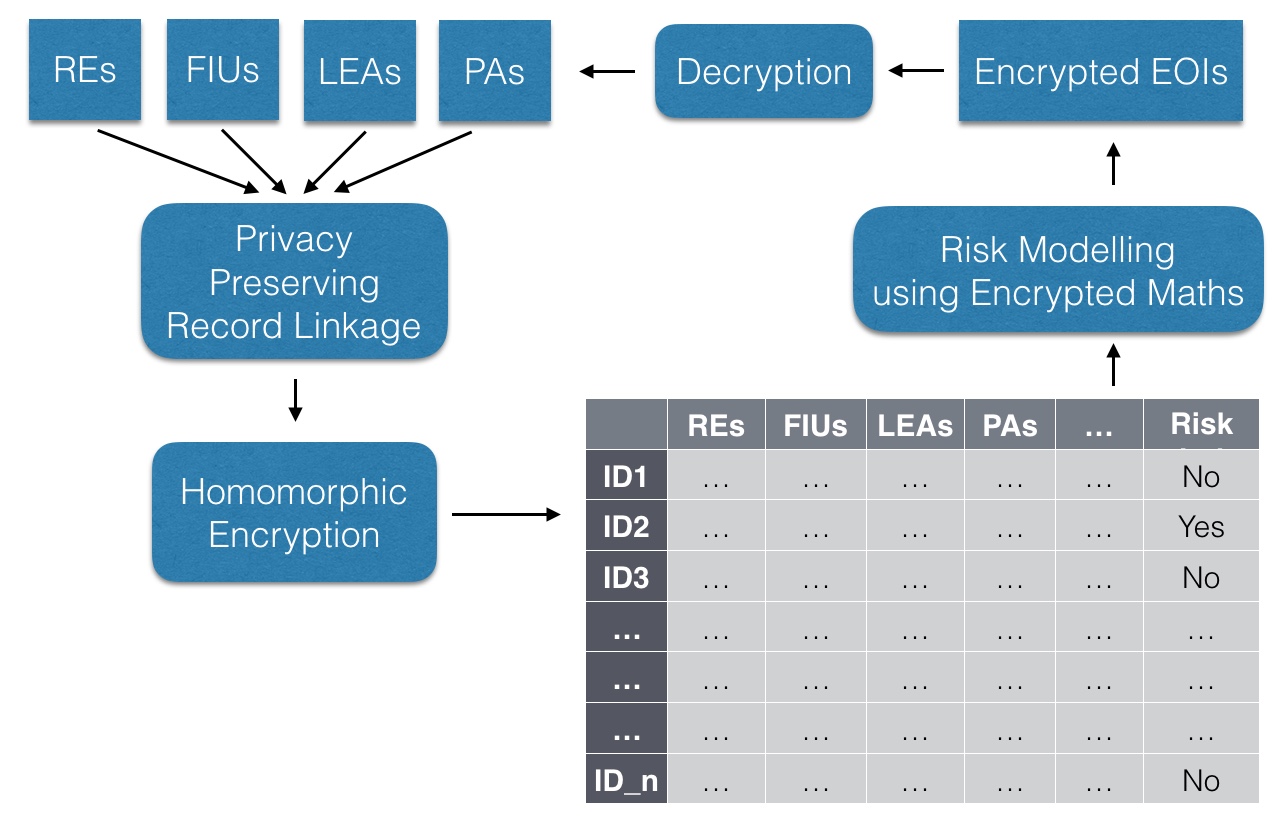}
    \caption{Privacy-enhanced version of Figure \ref{fig:fc-current}. 
    LEAs = Law Enforcement Agencies; 
    PAs = Partner Agencies; EOI = Entities of Interest}
    \label{fig:fc-future}
\end{figure}

To overcome the key limitations of the current state of financial intelligence, Figure \ref{fig:fc-future} outlines a potential solution that leverages distributed confidential computing technologies.
The assumption is that FIUs, reporting entities and law-enforcement agencies are willing to collaborate in good faith without requiring sweeping changes to AML/CTF legislation globally.

In the proposed approach, rather than traditional entity resolution, data from all participating organisations are first matched using privacy-preserving record linkage (PPRL) algorithms \cite{christenRS20}. These algorithms identify records across different databases that correspond to the same entity in a manner that safeguards privacy, ensuring no plaintext values are disclosed to any party involved.

Once the records are matched, the data are encrypted using either homomorphic encryption (Section \ref{subsec:HE}) or secure multiparty computation schemes (Section~\ref{subsec:mpc}). These encryption methods allow mathematical operations to be performed directly on encrypted data, resulting in a table similar to those discussed earlier but with two key differences:
\begin{itemize}\itemsep1mm\parskip0mm
 \item All values in the table are encrypted, and the columns are distributed across multiple databases managed by different agencies.
 \item AI and machine learning algorithms for detecting risks and suspicious entities can still be run on the encrypted data using homomorphic encryption or secure multiparty computation. 
 When a small set of entities of interest is identified, their data can be decrypted without compromising the privacy of the broader population.
\end{itemize} 
Distributing encrypted data across multiple organisations and databases does introduce some complexity to AML/CTF risk-modelling workflows, but the approach offers a pathway for collaboration while addressing privacy concerns and ensuring compliance with existing legal frameworks.

To help us navigate the opportunities offered by privacy technologies for financial intelligence, it is essential to understand that there are three fundamental computational problems to address in AML/CTF risk modelling, in increasing order of technical complexity:
\begin{enumerate}\itemsep1mm\parskip0mm
 \item Private Intersection Identification: How can two FIUs, each maintaining its own list of suspicious entities, securely exchange overlapping data points without exposing their full datasets with each other? How might this process scale across multiple FIUs?
 \item Cross-Institutional Pattern Detection: Given a criminal typology defined as a graph pattern (where nodes represent entities and edges denote relationships), how can we identify all entities matching this pattern when the underlying graph pattern spans multiple financial institutions?
 \item Federated Model Learning: How can supervised or unsupervised machine learning models be trained on behavioural and transactional data distributed across multiple financial institutions, where no single entity holds complete datasets?
\end{enumerate}
These three computational problems are distilled from the authors' experience working both directly and indirectly with Australia's financial intelligence and banking communities over an extended period of time, and backed up by literature review.

In the rest of this paper, we will first survey some key privacy technologies in \S~\ref{sec:privacy tech} before providing in \S~\ref{sec:pp-info-sharing} illustrative examples and detailed pointers to how such privacy technologies can be used to address the three computational problems listed above.
Solutions to these three computational problems can be, and are often, used as building blocks to develop more complete systems to address challenging real-world financial intelligence problems, some of which we survey in \S~\ref{sec:case studies}.

\section{Privacy Technologies}\label{sec:privacy tech}
In the above, we highlighted a central challenge in multiparty financial intelligence, how can multiple parties work together effectively when none of them can be assumed to be fully trusted? We also noted that privacy technologies can serve as a technological means to enable such collaboration, allowing multiple parties to jointly combat financial crime under specific cooperation models. We refer to this form of multiparty financial intelligence conducted under explicit data privacy and security constraints as privacy-preserving financial intelligence.

Privacy-preserving financial intelligence should be understood as a system level concept rather than a single method. Different privacy technologies play different roles within this system, but they share a common objective of enabling analysis and modelling, while maintaining data security and privacy guarantees.

This section begins with secure multiparty computation (MPC), as it is one of the most widely adopted mechanisms for protecting data in collaborative settings. MPC enables multiple parties to jointly perform computations without revealing their private inputs. Building on this foundation, we introduce homomorphic encryption (HE), which supports MPC by enabling computation over encrypted data and is therefore often used as a cryptographic primitive in MPC designs. We then move beyond MPC centric settings and discuss several problem scenarios where MPC is not the primary solution. For these cases, we introduce privacy technologies that are widely used in practice, namely hashing, federated learning (FL), and differential privacy (DP), each addressing privacy risks from a different perspective.

\subsection{Secure Multiparty Computation}\label{subsec:mpc}

The goal of secure multiparty computation (MPC) is to enable multiple parties to jointly evaluate a pre-agreed function without revealing their individual private inputs. At its core, MPC removes the need to trust the computation process itself. Under an MPC protocol, each participant learns only the final output (or an authorised share of it) and cannot infer additional information about other parties' inputs beyond what is inherently revealed by that output. In doing so, MPC shifts cross-institutional collaboration away from data centralisation or reliance on a trusted third party, and instead enforces privacy and correctness directly at the protocol level through cryptographic mechanisms.

More generally, MPC can be viewed as a general-purpose framework for privacy-preserving computation. Whenever a collaborative task can be formalised as a function, MPC provides a mechanism for jointly evaluating that function without requiring participants to share their raw data.

Because MPC is a strict security framework, its practical deployment depends on clearly defined security guarantees. In particular, privacy ensures that an adversary learns nothing about an honest party's input beyond what follows from the output itself. Correctness guarantees that the protocol produces the correct result, even in the presence of adversarial behaviour. Stronger notions may also be required in certain settings: guaranteed output (or robustness) ensures that adversaries cannot prevent honest parties from obtaining the result by aborting the protocol, while fairness requires that an adversary receives the output if and only if the honest parties do.

In addition to these properties, MPC protocols are analysed under explicit adversarial models. In the semi-honest (honest-but-curious) model, adversaries follow the protocol specification but attempt to extract additional information from the execution. In the malicious model, adversaries may deviate arbitrarily from the protocol, for example by sending malformed messages, aborting prematurely, or manipulating inputs and outputs. The assumed adversarial model directly influences both the strength of the security guarantees and the complexity of the resulting protocol.

In the context of financial intelligence, these capabilities can be illustrated through a common investigative scenario (discussed further in \Cref{subsec:spl}). Different financial institutions (FIs) maintain their own customer lists, while a financial intelligence unit (FIU) may need to determine which entities on its watchlist have relationships with which banks. Here, the FIU and the banks act as participating parties, their private inputs are their respective customer lists, and the agreed computation is the intersection between the FIU's list and those of the banks. MPC enables this joint analysis to be conducted without exposing any party's complete customer list.

\subsection{Homomorphic Encryption}
\label{subsec:HE}

Within the MPC framework, privacy protection is not achieved by simply hiding the output, but by ensuring that computation itself can be performed without exposing sensitive inputs. For this reason, MPC protocols typically rely on multiple cryptographic mechanisms rather than a single technique. Homomorphic encryption (HE) is one such foundational capability.

The defining feature of HE is that it allows certain computations to be performed directly on encrypted data, producing results that, once decrypted, are equivalent to those obtained from computations on plaintext. In MPC settings, this property enables parts of a computation to be executed without revealing intermediate values, thereby reducing direct data exposure between participants.
\begin{equation}
\label{eq:he}
\text{Dec}(\text{secret key}, \text{Enc}(\text{public key}, m_1) \circ \text{Enc}(\text{public key}, m_2)) = m_1 \diamond m_2,
\end{equation}

From this perspective, HE is not a replacement for MPC, but rather one of its supporting building blocks. In practice, HE is often combined with other mechanisms, such as secret sharing or oblivious transfer, to securely encapsulate specific computation steps while balancing privacy guarantees, communication overhead, and system complexity. At the same time, HE can also be used independently of a full MPC framework. In scenarios where the computation structure is relatively simple, for example, when one party holds the data and another performs the computation, HE allows encrypted data to be processed externally without revealing the underlying inputs. This enables analysis or evaluation without requiring raw data sharing.

HE schemes are commonly classified according to the types of operations they support and the number of operations that can be performed before decryption becomes unreliable. The most general form is Fully Homomorphic Encryption (FHE), which supports an unbounded number of operations on encrypted data. The practical feasibility of FHE was established following Gentry's breakthrough construction \cite{gentry2009fully}, which sparked significant research interest, particularly in lattice-based cryptography, a field now widely regarded as central to quantum-resistant encryption \cite{alagic2022status}.

It is also important to note that HE schemes are inherently probabilistic, which means encrypting the same plaintext multiple times produces different ciphertexts. This property ensures semantic security, which prevents adversaries from inferring information about the plaintext from ciphertext patterns.

In financial intelligence applications, such standalone use of HE may arise in outsourced computation of sensitive transaction features, risk metrics, or aggregate statistics. Financial institutions can retain full control over their data while leveraging external compute resources or cross-organisational collaboration, without disclosing plaintext information.

Despite its strong privacy guarantees, the applicability of HE remains constrained by performance considerations, supported computation types, and overall system complexity. As such, HE is best understood either as a specialised independent capability within the privacy-preserving toolkit, or as a key component within more complex MPC-based solutions.

\subsection{Hashing Techniques}
\label{sec:hashing}

In cross-institutional data collaboration, a fundamental but often overlooked problem is how to preserve matchability between entities or relationships while avoiding direct exposure of sensitive identifiers. At this level, hashing provides a lightweight yet practical solution.
Hashing works by applying a one-way mapping to original identifiers such as customer IDs, transforming them into fixed length representations. For example, 
\begin{align*}
    \text{Hash: Hello} \rightarrow \text{2CF24DBA5FB0A30E26E83B2AC5B9E}
\end{align*}
While the original identifiers cannot be recovered, consistent hashing allows entities or relationships across different datasets to be matched. In other words, hashing does not hide whether two records refer to the same entity, it hides who that entity is.

From a functional standpoint, hashing is not a joint computation mechanism and does not attempt to provide comprehensive protocol level privacy guarantees. Its primary value lies in reducing direct exposure of sensitive identifiers during data exchange and comparison, thereby offering a baseline level of privacy protection. This makes hashing widely used in practice for identity alignment, relationship linking, and preliminary data filtering.

In financial intelligence contexts, hashing is commonly applied as a preprocessing step for customer or account identifiers. This allows financial institutions and regulators or investigators to assess potential overlaps or business relationships without sharing plaintext identity information. While such approaches do not eliminate all privacy risks, they can significantly reduce unnecessary data exposure when computation goals are well defined and trust boundaries are controlled.

As suggested by Chi et al.~\cite{chi2017}, properties such as uniform distribution of hash outputs, consistency of mapping, computational efficiency, and appropriate handling of input and output domains are important when selecting a hash function for practical applications. More broadly, the effectiveness of hashing depends on several underlying assumptions, including the size of the input space, the consistency of hashing rules across participants, and resilience against auxiliary information attacks. In higher-risk or more complex multiparty collaboration settings, hashing alone may be insufficient; it is therefore typically combined with stronger privacy-preserving mechanisms rather than deployed as a standalone solution.

\subsection{Federated Learning}

In some collaborative settings, privacy risks arise not from a single joint computation, but from long term, iterative model training processes. In such cases, participants seek to collaboratively build and update models over time while keeping their data within local domains, rather than jointly evaluating a fixed function. Federated learning (FL) is designed for precisely this type of collaboration.

\begin{figure}
\centering
\includegraphics[width=0.75\textwidth]{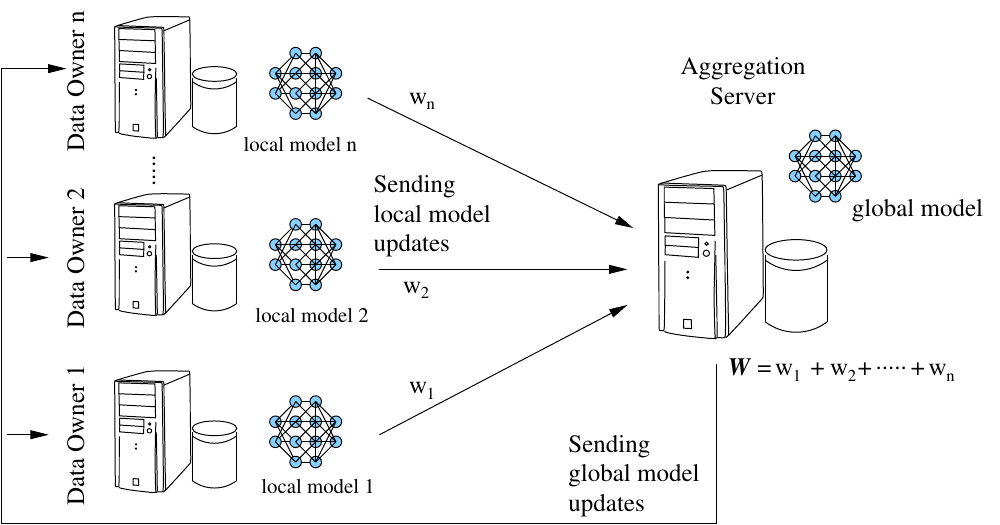}
\caption{An example federated learning (FL) architecture with $n$ data 
owners. In the setting, data owners only exchange the model gradient ($w$) 
with the aggregation server. In each training iteration, each data owner $i$
trains its own model with their local data and send the corresponding local model
updates $w_i$ to the aggregation server. After aggregating all
the received local model updates, the aggregations 
server returns the new global model updates ($W$)
to each data owner.}
\label{fig:fl-overview}
\end{figure}

The core idea of federated learning is that data remains local to each participant, model training is performed independently on local datasets, and only model updates are shared with a coordinating entity for aggregation into a global model as shown in \Cref{fig:fl-overview}. From a collaboration perspective, FL does not require sharing raw data or encrypting an entire joint computation. Instead, it follows a local computation and global aggregation paradigm.

This leads to a fundamental structural difference between FL and MPC. While MPC focuses on securely evaluating a predefined function, FL is concerned with the continuous evolution of a model over time. In large scale, long running, and frequently updated training tasks, attempting to replicate FL using MPC would be impractical due to excessive system complexity, communication overhead, and reduced training efficiency.

It should be noted, however, that although FL avoids direct data sharing, privacy risks do not disappear entirely. Model parameters or gradient updates may, under certain conditions, leak statistical properties or even individual level information about local datasets \cite{aledhari2020,nguyen2021,zihao2022}. Consequently, FL is often combined with additional privacy technologies, such as homomorphic encryption or differential privacy, to mitigate information leakage during the training process. 

Within financial intelligence, FL is particularly well suited to scenarios where multiple institutions wish to jointly train models for risk detection, anomaly identification, or behavioural analysis without centralising sensitive data. Participants retain control over their data while benefiting from improved model performance through collaboration. In such settings, the key challenge is not how to securely compute a single result, but how to balance model utility and privacy risk over sustained collaboration, a challenge for which FL offers distinct advantages over MPC.

\subsection{Differential Privacy}

Across the privacy technologies discussed so far, the primary focus has been on protecting input data and ensuring trustworthy computation. However, in many real world applications, the results themselves can still leak sensitive information, even when inputs and computation processes are fully protected. Differential privacy (DP) addresses this class of risks directly.

The core principle of DP is to introduce carefully calibrated randomness into statistical outputs or model results, ensuring that the presence or absence of any single individual's data does not have a distinguishable impact on the output. 

\begin{definition}
\label{def:dp}
A randomised algorithm $\mathcal{A}$ is $\epsilon$-differentially 
private if for all 
databases 
$\mathbf{D}$ and $\mathbf{D'}$ that differ in at most one record, 
and for all sets $\mathcal{O}$ outputs, we have
\begin{equation*}
    Pr[\mathcal{A}(\mathbf{D})\in\mathcal{O}]\le
    \exp(\epsilon)\cdot Pr[\mathcal{A}(\mathbf{D'})\in\mathcal{O}],
\end{equation*}
where $Pr[\cdot]$ denotes the probability of an event.\\[-0.7em]
\end{definition}
Rather than hiding raw data or computation steps, DP explicitly constrains how much information can be inferred from released results.

This positioning fundamentally distinguishes DP from MPC. Even in a fully trusted joint computation environment, repeated queries or downstream inference based on exact outputs can gradually expose individual level information. DP mitigates this risk by limiting the sensitivity of outputs to individual data points, something that MPC, HE, or FL cannot achieve on their own.

In financial intelligence applications, DP is commonly used when publishing aggregate statistics, analytical reports, or model outputs. For example, when sharing risk distributions, trend analyses, or group level behavioural insights, DP mechanisms can preserve the overall analytical value while substantially reducing the risk of re-identifying individual customers or accounts. In these contexts, the primary privacy concern is not who performed the computation, but what can be inferred once the results are released or reused \cite{nissim2021privacy,li21}.

It is important to emphasise that differential privacy is not a zero cost solution. Introducing randomness to limit information leakage necessarily involves a trade-off between privacy guarantees and result accuracy. As such, DP is best understood as an output level privacy governance mechanism, complementing rather than replacing input level and computation level privacy technologies.

\section{Privacy-Preserving Information Sharing}
\label{sec:pp-info-sharing}

In this section, we revisit the three computational problems listed at the end of \S~\ref{sec:fintel} and further contextualise them using real-world financial intelligence scenarios wherein 
information-sharing between different parties is crucial to support effective detection of financial crimes.
In each scenario, we discuss the privacy risks
during multiparty collaborations and examine the feasibility of 
applying various privacy-preserving techniques to address these 
privacy risks.

It is important to recognise that each scenario may entail varying 
levels of privacy risks. These risks depend on the degree of 
trust among different parties. The level of trust is basically inversely
proportional to the risks of sharing information that have to be shouldered by the collaborating parties, which is in turn a function of the prevailing government
legislation (or lack thereof) and applicable penalties in the event of a privacy breach.

The parties involved in conducting financial intelligence analysis
today may include both public sectors (domestic or international) 
and private sectors. Generally speaking, information-sharing between
different domestic public sectors involves a high level of trust. 
In contrast, information sharing between public and private sectors, or among
private sectors, often entails a low level of trust. This lack of trust may stem
from various factors, including the absence of clear legislation and threats of business 
competition. For instance, financial institutions 
tend to refrain from sharing customer transaction data among themselves, 
despite the potential benefits such information-sharing could offer 
in enhancing their anti-money laundering controls. 
Privacy technologies provide the greatest value in low-trust settings
that exist either because the parties involved cannot trust each other 
or a third party for commercial/legal reasons, or when there is 
ambiguity on whether a public-good sharing of information crosses
the threshold into unacceptable privacy intrusion.

\subsection{Suspicious Persons List}\label{subsec:spl}
\paragraph*{Problem statement} 
Suppose an FIU possesses a moderate-sized list of suspicious individuals (or, more generally, commercial entities) obtained, say, from its own investigations and law-enforcement agencies and the FIU wishes to acquire additional information about those suspicious individuals, including where they conduct their financial activities and whether they are known to other FIUs.
This can be done if the FIU can safely distribute the Suspicious Person List (SPL) to other FIUs and financial institutions to verify the association of individuals on the list with specific financial entities before engaging in further targeted information exchanges with those financial entities.

\paragraph*{Privacy risks} If the SPL is shared with the financial 
institution in a manner that allows for the identification of individuals
on the list, there arise the following privacy concerns, especially 
in the low-trust public-to-private information-sharing setting. 

\begin{itemize}\itemsep1mm\parskip0mm
    \item (Accidental Tip-off) The financial institution may 
    inadvertently disclose the FIU's suspicions to an existing customer,
    prompting the customer to take additional measures to conceal
    their activities. 
    
    \item (Debanking and Displacement) The financial institution may 
    deliberately cease its banking relationship with a suspicious individual before 
    conclusive evidence can be gathered to substantiate or refute the
    individual's suspected wrongdoing. This action may be deemed 
    discriminatory if the financial institution's decision to terminate
    services is also influenced by personal factors such as the 
    individual's nationality, race, or name. Further, suspected 
    individuals that are debanked will also move on to other parts 
    of the financial system, including the shadow financial system, 
    and the FIU will lose sight of them.

    \item (Need to Know) The FIU can inadvertently disclose the 
    identities of suspicious entities who are not customers of a 
    financial institution, and that would be an unacceptable
    violation of those individual's privacy and the need-to-know 
    principle.
    
\end{itemize}

\paragraph*{Proposed solution}

The problem described above of identifying common individuals between 
the SPL and each participating financial institution's customer database
can be formulated as a private set intersection (PSI) problem. 
The goal of PSI is to identify the 
common elements that appear in the parties' sets and nothing else. 

We now describe a simple semi-honest PSI protocol from \cite{freedman2004efficient} that utilises
polynomial evaluation and homomorphic encryption to ensure the protection of unmatched individuals in the SPL.
For detailed security and 
efficiency analyses, readers are encouraged to refer to 
\cite{freedman2004efficient}. For simplicity, it is assumed that each 
individual can be uniquely identified by an identification (ID) number, 
which can be derived from a national ID (in some countries) or a 
combination of personal information.
Here are the steps of the protocol.
\begin{enumerate}\itemsep1mm\parskip0mm
    \item Assuming the FIU's list of suspicious individuals is 
    $S = \{s_1, \dots, s_n\}$, the FIU first defines the polynomial 
    \begin{align*}
        f(x) = (x-s_1)(x-s_2) \cdots (x-s_n), 
    \end{align*}
    whose roots are exactly the elements in the set $S$. The
    question of whether an arbitrary $s'$ is in $S$ can be checked by evaluating $f(s')$. 
    
    \item 
    The FIU uses polynomial interpolation to find the coefficient 
    representation of $f(x)$    
    \begin{align*}
        f(x) = a_0 + a_1 x + \cdots + a_n x^n,
    \end{align*}
    which is uniquely defined according to the Fundamental Theorem of Algebra.
    The FIU then encrypts the coefficients of $f(x)$ using HE and sends 
    the encrypted coefficients and the public key to the other parties:
    \begin{align*}
        \text{Enc}(f) = \{\text{Enc}(a_0), \dots, \text{Enc}(a_n)\}.
    \end{align*}
    
    \item Upon receiving the encrypted $f$, each financial institution evaluates the 
    encrypted polynomial $\text{Enc}(f)$ at each element $d \in D$ 
    in its customer database, then multiplies the evaluation result by a 
    random integer $r$ and adds it to $d$. The property of HE guarantees 
    that the result of these operations is equivalent to the following 
    \begin{align*}
        \text{Enc}(r f(d) + d), \forall d \in D.
    \end{align*}
    The above formula ensures that if $d$ is a positive
    match then FIU is able to identify it from the output, and if 
    $d$ is not a positive match then FIU learns nothing about $d$ 
    from the output. 

    \item Once the matching outputs are shared, FIU decrypts the 
    outputs and finds those elements that also appear in its suspicious
    person list.
    
\end{enumerate}

By executing this PSI protocol, the FIU and financial institutions 
collaborate in a confidential manner, enabling the FIU to identify 
individuals who appear on both the FIU's SPL and the databases of 
participating financial institutions. This facilitates the FIU in 
requesting additional information from appropriate financial 
institutions to conduct further investigations. Simultaneously, 
individuals listed on the SPL but absent from a financial 
institution's database remain undisclosed to that institution.
Similarly, individuals present in a financial institution's database
but not flagged on the SPL remain confidential to the FIU.
  
\paragraph*{Industrial grade solutions}
The FIU.net operated jointly by FIUs in the European Union has 
offered privacy-preserving sharing of suspicious entities for over 
a decade now, using the Ma3tch algorithm \cite{balboni2013privacy}.
A recent legal review of FIU.net can be found in 
\cite{mouzakiti2020cooperation}.
In Australia, the Australian Financial Crime 
Exchange (\url{http://afcx.com.au}) is a platform for
financial institutions to share lists of entities of interest, 
from a financial crime and cyber crime perspective, with each other 
and, selectively, with the Australian Taxation Office.

There are now many fast PSI algorithms that can operate at large 
scale, including \cite{chen2017fast, rosulek2021compact, chase2020private}.
A key issue with PSI is that an adversary can use repeated PSI 
computations to completely recover an organisation's entire dataset.
Proposals like \cite{purcell2023split} seek to address this issue by 
introducing DP protection into PSI computations.

\subsection{Transaction Flow Tracing and Typology Matching}\label{subsec:typology matching} 

\paragraph*{Problem statement} 
A key problem in financial crime investigation is tracing 
the entire digital footprint of ``dirty money'' within the financial system, allowing investigators 
to clearly track the movement of funds across different financial institutions, how they are processed and transferred and, crucially, their ultimate usage or destination.
We call this problem transaction-flow tracing.

A key task in financial intelligence supported by transaction-flow 
tracing is the identification of accounts that match certain 
money-laundering or terrorism-financing (ML/TF) typologies. 
The NDIS fraud described in \cite{brand2023fintracer} serves as an example.
Australia's National Disability Insurance Scheme (NDIS) provides financial support to 
eligible participants to improve their quality of life. However, 
continuous fraudulent activities against the NDIS scheme have resulted
in several billions of dollars in losses from its intended 
use \cite{ndis2020}. These fraudulent transactions share common 
patterns, crossing multiple organisations and eventually being 
transferred overseas. No single institution in the financial system
would have all the data required to identify accounts and entities 
that exhibit such behaviour.
We will show in this section how privacy technologies can help tackle this broad class of problems.

\paragraph*{Privacy risks} 
This scenario exemplifies a case of government-to-private and 
private-to-private information sharing, which are generally low-trust settings.
Complex ML/TF typologies can usually be described as a graph pattern,
where nodes correspond to accounts/entities and edges correspond to 
transactions between those nodes and there can be non-trivial conditions attached
to the nodes and edges. To enable the identification of entities that 
match a complex ML/TF typology that span the financial system, we would 
need accounts and transaction data from multiple financial institutions 
to be matched at scale before a filtering or sub-graph matching operation
is performed. This is obviously highly intrusive to entities who are not
involved in financial crimes, so we need a way of performing the 
large-scale data-matching and graph pattern searching in a 
privacy-preserving manner.

\paragraph*{Proposed solution} 
The FinTracer algorithm described in \cite{brand2023fintracer} is a computationally efficient way to conduct breadth-first
search on large distributed graphs in homomorphically-encrypted 
space. 
The algorithm enables Financial Intelligence Units to start with some known suspicious entities and then work with
law-enforcement agencies and financial institutions to find the entire footprints of those entities across the financial system 
\begin{enumerate}\itemsep1mm\parskip0mm
    \item without having to disclose the actual identities of the suspicious entities to financial 
    institutions, thus avoiding the risk of accidentally tipping off 
    the entities; and
    \item without having to compromise the privacy of unrelated entities even 
    though their data need to be processed.
\end{enumerate}
The key insight behind the FinTracer algorithm is that large distributed graphs can be represented as sparse matrices and the breadth-first search algorithm can be expressed as a sequence of matrix multiplication operations \cite{burkhardt2021optimal}. 
The FinTracer algorithm can be used as a foundational routine to detect a 
wide range of financial crime typologies (i.e. graph patterns).
Its adoption can thus meaningfully advance the state-of-the-art in financial intelligence 
agency's ability to detect complex financial crimes that span multiple organisations in a privacy-preserving manner at scale.

\begin{figure}
    \centering
    \resizebox{0.5\textwidth}{!}{
\begin{tikzpicture}[
    thick,
    every node/.style={draw,circle},
    fsnode/.style={},
    ssnode/.style={},
    every fit/.style={
        rectangle,
        draw,
        rounded corners=6pt,
        inner sep=6pt
    },
    ->,
    shorten >= 3pt,
    shorten <= 3pt
]

\begin{scope}[start chain=going below,node distance=3mm]
\foreach \i in {1,2,3,4}
  \node[fsnode,on chain] (a\i) {a\i};
\node[fsnode,fill=orange] at (a1) {a1};
\end{scope}

\begin{scope}[xshift=3cm,yshift=0.5cm,start chain=going below,node distance=3mm]
\foreach \i in {1,2,...,5}
  \node[ssnode,on chain] (b\i) {b\i};
\end{scope}

\begin{scope}[xshift=6cm,yshift=0.5cm,start chain=going below,node distance=3mm]
\foreach \i in {1,2,...,5}
  \node[ssnode,on chain] (c\i) {c\i};
\end{scope}

\begin{scope}[xshift=9cm,start chain=going below,node distance=3mm]
\foreach \i in {1,2,...,4}
  \node[ssnode,on chain] (d\i) {d\i};
\end{scope}

\node[fit=(a1)(a4),label=above:$A$] {};
\node[fit=(b1)(b5),label=above:$B$] {};
\node[fit=(c1)(c5),label=above:$C$] {};
\node[fit=(d1)(d4),label=above:$D$] {};

\draw (a1) -- (b1);
\draw (a1) -- (b2);
\draw (b1) -- (c1);
\draw (b2) -- (c2);
\draw (c1) -- (d1);
\draw (c2) -- (d1);

\end{tikzpicture}
}
    \caption{A simplified transaction network to demonstrate
    FinTracer. Accounts with cross-organisation
    transactions are linked by directed edges. There are four 
    financial institutions A, B, C and D, each having a number 
    of accounts. The account $a_1$ in A directly received NDIS payment.}
    \label{fig:fintracer}
\end{figure}

To see how this works, suppose AUSTRAC is interested in 
identifying the accounts (from multiple financial institutions) 
that transferred NDIS funds overseas. These accounts either received
funds directly from NDIS or indirectly via a limited number of 
cross-institution transactions. 
The FinTracer tag $t_{\le}$ is implemented to keep track of accounts
with at most $k$ cross-institution transactions. 
These tags are homomorphically encrypted to securely follow
NDIS funds in the entire transaction network. 
The key components of the FinTracer process are the ElGamal encryption
scheme \cite{elgamal1985public} over the additive Curve25519 for its
performance efficiency \cite{bernstein2006curve25519}, and the 
aggregations of the tag values across financial institutions over 
multiple iterations of the breadth-first search algorithm.

\Cref{fig:fintracer} is a simplified example of the NDIS fraud 
scenario. In this example, there are four financial institutions. 
The only account that directly received suspicious NDIS funds is a1 in institution A.
All institutions are only aware of the source and destination accounts that are involved in 
the transactions they process. In its simplest form 
(Algorithms 1 and 2 \cite{brand2023fintracer}), the FinTracer protocol
does the following steps:

\begin{itemize}\itemsep1mm\parskip0mm
    \item Initialisation: Each financial institution sets the initial 
    FinTracer tag of each of its customer account $x$ to either $\text{Enc}(1)$ or 
    $\text{Enc}(0)$, depending on whether $x$  receive NDIS funds directly. 
    In the example above, 
    \begin{itemize}
        \item $t_{\le}(a1)=\text{Enc}(1)$ and $t_{\le}(\{b1, b2, c1, c2, d1\})=\text{Enc}(0)$ (the set notation denotes that the tag is set for element in the set).
        \item The other accounts do not have cross-institution
        transactions, hence are out of considerations by the protocol.
    \end{itemize}

    \item Iteration 1: In the first iteration, the initialised tags are 
    passed to the neighbouring institutions following the transaction flow.
    The receiving institutions then aggregate the tags with the received
    tags. In the example, the shared partial mappings are 
    \begin{itemize}
        \item A $\rightarrow$ B: $P^A_B((a1,b1))=\text{Enc}(1), P^A_B((a1,b2))=\text{Enc}(1)$.
        \item B $\rightarrow$ C: $P^B_C((b1,c1))=\text{Enc}(0), P^B_C((b2,c2))=\text{Enc}(0)$.
        \item C $\rightarrow$ D: $P^C_D((c1,d1))=\text{Enc}(0), P^C_D((c2,d1))=\text{Enc}(0)$.
    \end{itemize}
     After receiving these partial mappings, the aggregated tags are 
     \begin{itemize}
        \item $t_{\le}(\{a1,b1,b2\})=\text{Enc}(1)$ and $t_{\le}(\{c1, c2, d1\})=\text{Enc}(0)$
    \end{itemize}
    since $t_{\le}(b1) = t_{\le}(b1) + P^A_B((a1,b1)) = \text{Enc}(0) + \text{Enc}(1) = \text{Enc}(1)$ and similarly for $t_{\le}(b2)$.
     The tag values of $c1$, $c2$, $d1$ remain $\text{Enc}(0)$ since the 
     initial and received tags are $\text{Enc}(0)$.

    \item Iteration 2: The partial mappings are updated by institution B yielding 
    \begin{itemize}
        \item A $\rightarrow$ B: $P^A_B((a1,b1))=\text{Enc}(1), P^A_B((a1,b2))=\text{Enc}(1)$.
        \item B $\rightarrow$ C: $P^B_C((b1,c1))=\text{Enc}(1), P^B_C((b2,c2))=\text{Enc}(1)$.
        \item C $\rightarrow$ D: $P^C_D((c1,d1))=\text{Enc}(0), P^C_D((c2,d1))=\text{Enc}(0)$.
    \end{itemize}
     After receiving these partial mappings, the aggregated tags are 
     \begin{itemize}
        \item $t_{\le}(a1)=\text{Enc}(1)$, 
        $t_{\le}(\{b1, b2\})=\text{Enc}(2)$, 
        $t_{\le}(\{c1, c2\})=\text{Enc}(1)$, 
        $t_{\le}(d1)=\text{Enc}(0)$.
    \end{itemize}
    
    \item Iteration 3: The partial mappings are updated by B and C yielding
    \begin{itemize}
        \item A $\rightarrow$ B: $P^A_B((a1,b1))=\text{Enc}(1), P^A_B((a1,b2))=\text{Enc}(1)$.
        \item B $\rightarrow$ C: $P^B_C((b1,c1))=\text{Enc}(2), P^B_C((b2,c2))=\text{Enc}(2)$.
        \item C $\rightarrow$ D: $P^C_D((c1,d1))=\text{Enc}(1), P^C_D((c2,d1))=\text{Enc}(1)$.
    \end{itemize}
     After receiving these partial mappings, the aggregated tags are 
     \begin{itemize}
        \item $t_{\le}(a1)=\text{Enc}(1)$, 
        $t_{\le}(\{b1, b2\})=\text{Enc}(3)$, 
        $t_{\le}(\{c1, c2\})=\text{Enc}(3)$, 
        $t_{\le}(d1)=\text{Enc}(2)$.
    \end{itemize}
\end{itemize}
The FinTracer tags are now sent to the FIU and decrypted. Only those accounts with a non-zero tag have received money that can be traced either directly or indirectly to suspicious NDIS payment.

\paragraph*{Finding key players in transaction networks}
Once the small number of accounts and entities that matched a certain
ML/TF typology is detected using a technology like FinTracer, an FIU 
like AUSTRAC has the legal power to request much more comprehensive 
information about those accounts and entities from financial
institutions because the request for information or notice to produce
information is now targeted and backed by genuine grounds for suspicion.
(The remaining small-scale privacy concerns can mostly be addressed using differentially private graph search algorithms like \cite{kearns2016private}.)
With such a dataset, it then makes sense to calculate connected components
and risk metrics like PageRank scores and betweenness centrality scores,
some of which have been shown to be useful features to enhance the 
accuracy or confidence of machine learning or rule-based financial 
intelligence classification models \cite{molloy2017graph,sangers2019secure,van2024privacy}.

\paragraph*{Industrial grade solutions}
An open-source, scalable implementation of FinTracer can be found at 
\url{http://github.com/AUSTRAC}.
The implementation includes a C/Cuda library for additive HE with 
ElGamal over the Ed25519 elliptic curve.
A number of algorithms for different graph operations are 
included in \url{http://github.com/AUSTRAC/ftillite/Documentation/Purgles}.
The extension of FinTracer, which allows only linear operations because
of the use of additive HE, to non-linear set and graph operations to 
allow the detection of more-or-less arbitrarily complex criminal 
typologies is described in \cite{brand2023nonlinear}.
Note that differential privacy noise is added in several steps in 
these algorithms to avoid variations of database-reconstruction attacks \cite{dinur2003revealing, dwork2007price}.

Techniques described in this section can also be extended to provide more general SQL-like queries on encrypted data, using ideas described in CryptDB \cite{popa2011cryptdb, yousuf2021systematic}.
In this area, Enveil and Duality provide industry-grade solutions with their products, Enveil ZeroReveal Search and Duality Query Engine. 
Such solutions allow organisations to run SQL-like queries on decentralised databases, ensuring data remains within the owner's environment and supporting compliance with data sovereignty laws. 
Additionally, these secure systems encrypt both query content and results, keeping them private and visible only to the querier.

\subsection{Collaborative Model Learning}

\paragraph*{Problem statement}

The use of supervised learning and anomaly detection algorithms to 
learn financial crime models from data has long been a topic of interest within
financial intelligence \cite{sangers2019secure,van2024privacy}, but the practical utility of such approaches
has been hampered by a lack of quality labelled data in the case of 
supervised learning algorithms, and high false positive rates in the 
case of anomaly detection algorithms.

Recent advances in privacy-preserving FL have the potential to alleviate
some of these issues by allowing different business units within a 
financial institution and even multiple financial institutions to 
jointly learn models from their collective data.
In this section, we describe some of the key ideas behind some exploratory work 
in this area.

\paragraph*{Privacy risks}

To enable FL, we would need accounts and transaction data from multiple
financial institutions to be matched, often probabilistically, at scale 
before a machine learning algorithm can be run.
In the case of supervised learning, the participating parties would 
also need to share, as labelled data, entities that are either known or
suspected to be engaging in financial crime.
This is obviously highly intrusive to entities who are not involved in 
financial crimes, and also causes the kind of tipping-off issues 
discussed in Section \ref{subsec:spl} for those (suspected to be) involved
in financial crimes.
All these imply that we need scalable and secure ways to perform
large-scale probabilistic data matching and model learning in a 
privacy-preserving manner.
The publication and reuse of machine learning models by multiple organisations also open the door to attacks like membership inference that need to be managed \cite{mothukuri2021survey}.

\paragraph*{Proposed solutions}

As discussed, the first step in FL is privacy-preserving
data matching. Unlike transaction flow tracing, where the source and 
target of each transaction are uniquely identifiable for exact matching, 
there can be inconsistencies in how different financial institutions 
collect and store individual information, such as names, addresses, 
driver's licences, business names, and other details. Consequently, 
data matching and entity resolution in federated learning - both within 
and across financial institutions - are often performed probabilistically.
A comprehensive description of key privacy-preserving 
data matching/record linkage algorithms, including those that are built on 
Bloom Filters can be found in \cite{christenRS20}.

\begin{figure}
    \centering
    \includegraphics[width=0.75\linewidth]{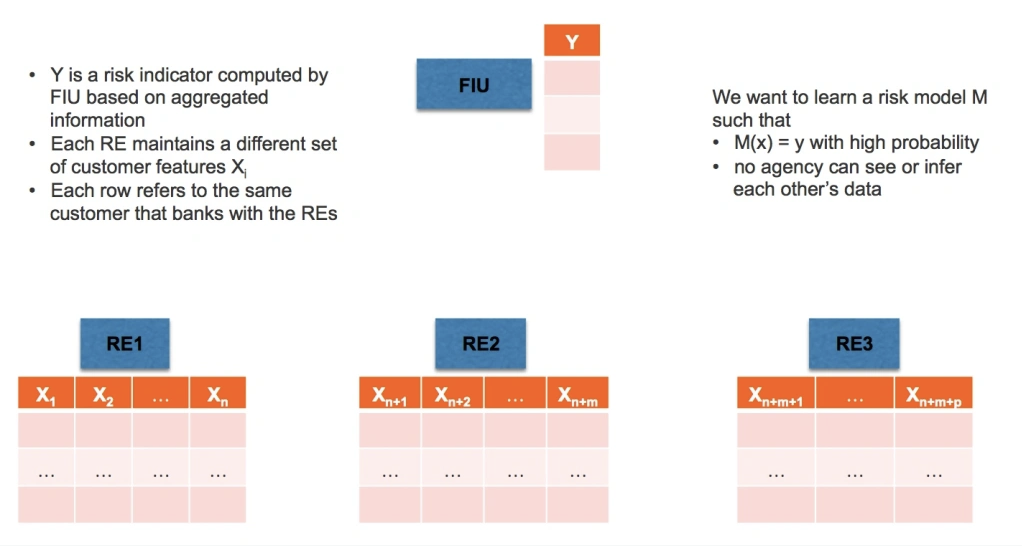}
    \caption{Federated learning setup for privacy-preserving risk modelling using linear regression}
    \label{fig:fl setup}
\end{figure}

We now describe a simplified example of federated learning to illustrate its applicability in financial intelligence.
The task is to learn a linear regression risk model using probabilistically matched data from multiple participating parties, while preserving privacy in a distributed manner. 
The target of the regression is the individual risk score, and the features for the regression 
are derived from all the data held by the REs. The setup is shown in Figure \ref{fig:fl setup} for three parties. 

More formally, given a training set $\{(\vb*{x}_l, y_l)\}_{l=1,\dots,m}$ 
that includes $m$ individuals, where the feature set 
$\vb*{x}_l = \vb*{x}_l^{(1)} \cup \cdots \cup \vb*{x}_l^{(p)}$ is the union
of the features from all parties, the task is to fit a linear regression
model 
    $\vb*{Y} = \vb*{X} \vb*{w} + \vb*{\epsilon}$
that predicts the target variable $\vb*{Y}$ as accurate as possible. 

The regression training is performed using gradient descent. The weight 
vector $\vb*{w}$ is initially assigned random values and then updated 
using the gradient of the loss function with respect to each weight 
element $w_j$ as follows, until convergence.
\begin{align}
\label{eq:lr_gradient}
w_j := w_j - \alpha \frac{1}{m} \sum_{l=1}^m (\langle \vb*{x}_l, \vb*{w} \rangle - y_l) x_{l,j} 
\end{align}
The algorithm converges to the optimal solution because the 
least-squares loss function is convex and the learning rate $\alpha$ 
controls the convergence rate. 

Below are the steps for the privacy-preserving distributed version 
of the algorithm that uses an additive HE system.
\begin{enumerate}\itemsep1mm\parskip0mm
    \item The FIU generates a public-private key pair ($pk$, $sk$) and
    shares the public key $pk$ with the REs.

    \item Each RE initialises its portion of the weight coefficients 
    $\vb*{w}^{(i)}$ randomly and computes a partial prediction for 
    each individual 
          \begin{align*}
              \vb*{Y}^{(1)} &= (\langle \vb*{x}^{(1)}_{1}, \vb*{w}^{(1)}\rangle, \langle \vb*{x}^{(1)}_{2}, \vb*{w}^{(1)} \rangle, \ldots, \langle \vb*{x}^{(1)}_{m}, \vb*{w}^{(1)} \rangle),\\
              \vb*{Y}^{(2)} &= (\langle \vb*{x}^{(2)}_{1}, \vb*{w}^{(2)}\rangle, \langle \vb*{x}^{(2)}_{2}, \vb*{w}^{(2)} \rangle, \ldots, \langle \vb*{x}^{(2)}_{m}, \vb*{w}^{(2)} \rangle),\\
              \vdots \\
              \vb*{Y}^{(p)} &= (\langle \vb*{x}^{(p)}_{1}, \vb*{w}^{(p)}\rangle, \langle \vb*{x}^{(p)}_{2}, \vb*{w}^{(p)} \rangle, \ldots, \langle \vb*{x}^{(p)}_{m}, \vb*{w}^{(p)} \rangle), 
          \end{align*}
          where $\vb*{x}^{(i)}_{l}$ is the feature set of individual 
          $l$ held by the $i^{th}$ RE.
          
    \item Each RE homomorphically encrypts its partial prediction 
    element-wise and participates in a round robin communication to 
    aggregate a complete prediction for each individual. The resulting
    encrypted prediction is then sent to the FIU for subsequent 
    computation. 
         \begin{equation*}
            \overline{\vb*{Y}} = \left(\sum_{i=1}^p \enc(pk, \vb*{Y}^{(i)}_1), \dots, \sum_{i=1}^p \enc(pk, \vb*{Y}^{(i)}_m)\right), 
         \end{equation*}
          
    \item The FIU decrypts the complete prediction and computes the
    difference from the actual values
            \begin{equation*}
                \vb*{E} = \dec(sk, \overline{\vb*{Y}}) - \vb*{Y},
            \end{equation*}
            and then sends the unencrypted $\vb*{E}$ back to the REs.
            
    \item Each RE then updates its portion of the weight vector 
    $\vb*{w}^{(i)}$ independently using \Cref{eq:lr_gradient}
            \begin{equation*}
                \vb*{w}^{(i)}_j = \vb*{w}^{(i)}_j - \alpha \frac{1}{m} \sum_{l=1}^m \vb*{E} \cdot x^{(i)}_{l,j}
            \end{equation*}
            where $j$ ranges over the variables held by the RE.

    \item Repeat steps 2 to 5 until the model has converged.         
\end{enumerate}

Throughout the entire computation, the FIU only sees the predicted
values from the REs, from which it cannot recover the data held by the
REs. Conversely, the REs only see the encrypted partial predictions
from each other and the gradients from the FIU, from which they 
cannot recover the data held by the FIU or other REs.

In the remainder of this subsection, we provide a survey of some innovative use of FL that holds promise for financial intelligence.
In the work of \cite{suzumura2022}, we see a reasonably compelling 
demonstration of how FL can significantly enhance
the detection capabilities of financial crime models. Through a 
series of case studies and rigorous experimental evaluations, this 
research highlights the dual achievement of improved detection 
performance and robust data privacy. 

Building upon this foundation, \cite{dhiman2023} introduces a 
novel approach that advances secure and collaborative model 
training across multiple financial institutions. By integrating
HE with FL, this approach achieves secure model aggregation and
parameter updates. The experimental results and evaluations 
presented reveal the approach's effectiveness in preserving data
privacy and security, thus fostering a collaborative environment
for financial data analysis across institutions.

Further extending this narrative, \cite{hardy2017} presents an 
end-to-end scheme designed for linear classifiers, employing 
additive HE to counter 
semi-honest adversaries 
in vertical FL scenarios. This research delves into the impact 
of entity resolution on learning performance, utilising a Taylor
approximation for the loss and gradient functions. The application
of Paillier additive HE within this context enables privacy-preserving
computations, showcasing an innovative method for handling
sensitive data securely.

Additionally, \cite{byrd2020} proposes a FL 
scheme tailored for fraud detection. In this scheme, individual
financial organisations enhance their model parameters with 
DP noise before transmitting them to a central server. This 
strategy effectively diminishes the server's ability to recover 
private information, advancing the cause of privacy in fraud 
detection efforts.

The above FL techniques collectively offer a glimpse into the forefront of 
financial intelligence. Their contributions highlight ongoing
advancements and set the stage for future innovations in the field.

\paragraph*{Industrial grade solutions}
There are several industrial-grade solutions like \cite{zhang17,zhang18} for conducting probabilistic matching of databases with millions to billions of entities 
and these can be used in conjunction with techniques like 
Bloom filters and HE to produce somewhat scalable privacy-preserving
probabilistic data-matching algorithms.
Privitar SecureLink\footnote{\href{https://docs.informatica.com/content/dam/source/resources/privitar-docs/data-privacy-platform/461/DPP-user/en/about-privitar-securelink.html}{SecureLink website}} offers a solution for securely joining data from various contributors without revealing sensitive identifiers or directly linking data across organisations. 

While federated learning is now a mature technology, progress in using federated learning at the industrial scale for financial intelligence applications is still somewhat hampered by general lack of labelled data to support the acquisition of robust  and trusted machine-learning models with low false-positive rates.
Having said that, several case studies and investigations have validated and implemented privacy-preserving solutions for specific scenarios. In particular, industrial solutions developed by Enveil, Duality, and Inpher have scaled effectively to address real-world challenges, with some initiatives leading to collaborations with other organisations to create domain specific products. The details are described in \Cref{sec:case studies}. 

\section{Real-World Case Studies}\label{sec:case studies}

In this section, we describe several case studies that have been trialled and, in some cases, deployed
in real-world applications. 
The presentation provides an update on some key projects first listed in \cite{ffis2021} and includes more recent additional work; the emphasis is on the business problem, participating financial institutions, and achieved business outcomes (when that is available.) 
The case studies vary in scope and complexity, with some focusing on a singular use case while others cover all the three major use cases described in \Cref{sec:pp-info-sharing}.
It is important to note that in the practical context, AML/CTF use 
cases can be affected by a large number of factors, including computational 
cost, operational cost, technical complexity, efficiency, data availability, 
quality and interoperability. 
These case studies are summarised in Table~\ref{tab:case-studies}.

\begin{table}
\centering
\caption{Overview of several case studies by different jurisdictions.}
\label{tab:case-studies}
\begin{footnotesize}
\begin{tabular}{|p{3.5cm}|p{5cm}|p{3cm}|}
\hline
\multicolumn{1}{|c|}{Project} &
  \multicolumn{1}{c|}{Participants} &
  \multicolumn{1}{c|}{Use case} \\ \hline \hline
Enveil ZeroReveal Inter-Bank Secure
and Private Data Collaboration~\cite{ffis2021} & An EU-based financial
institution and a third-party data provider 
& Transaction flow tracing and typology matching \\ \hline
Duality SecurePlus Query~\cite{ffis2021} & UK and Canadian Banks & Transaction flow tracing and typology matching \\ \hline
DANIE consortium~\cite{ffis2021} & UK-based consortium of banks & Suspicious persons list \\ \hline
Privitar SecureLink~\cite{ffis2021} & Public and private Institutes in an EU member state &  Suspicious persons list\\ \hline
BNY Mellon and Inpher XOR~\cite{InpherTechreport} & BNY Mellon & Collaborative learning\\ \hline
UK Tri-bank initiative~\cite{ffis2024} & Three UK banks & Transaction flow tracing\\ \hline
AUSTRAC's Fintel Alliance Alerting Project~\cite{chadderton2019,scott2024} & Led by AUSTRAC Australia and delivered by the public-private partnership Fintel Alliance&  Transaction flow tracing \\ \hline
AML Bridge Platform~\cite{fatf2022} & Ten banks and several non-banks in Estonia & Transaction flow tracing and typology matching\\ \hline
Transaction Monitoring Netherlands (TMNL)~\cite{van2024} & Five banks in Netherlands & Transaction flow tracing\\ \hline
DeepProtect~\cite{kanamori2022} & Five banks in Japan & Collaborative learning\\ \hline
Project Aurora~\cite{AuroraTechreport} & Led by Bank for International Settlements and participated by 29 financial institutions in six European countries& Collaborative learning \\ \hline
COSMIC~\cite{fatf2022,CosmicTechreport} &  Owned and operated by the Monetary Authority of Singapore (MAS). The is a co-development with the Commercial Affairs Department (CAD) of the Singapore Police Force, and six major banks (Citibank, DBS, HSBC, OCBC, Standard Chartered Bank and UOB).& Transaction flow tracing and typology matching \\ \hline
\end{tabular}%
\end{footnotesize}
\end{table}

\paragraph*{Enveil ZeroReveal Inter-Bank Private Collaboration~\cite{ffis2021}}

\noindent This is an initiative, started in 2019, by the UK Financial Conduct 
Authority (FCA) Global AML and Financial Crime TechSprint and Enveil
to develop a system that could enable data and knowledge sharing between 
stakeholders to identify and prevent money flows within and between 
criminal networks. This initiative was supported by several private 
UK banks including EY, BAE Systems, Refinitiv, HSBC, Barclays 
and ING. Their proposed system used homomorphic encryption to provide 
secure and automated multiparty collaboration which allows banks to 
query about higher risk customers. 
Since that FCA TechSprint, the Enveil suite of solutions have 
developed in maturity and are now used in multiple sensitive 
domains.\footnote{\href{https://www.enveil.com/enveil-wins-army-linchpin-contract-for-secure-ai/}{Enveil Wins Army Linchpin Contract to Deliver Secure AI}}

\paragraph*{Duality SecurePlus\textsuperscript{TM} Query on Financial Data~\cite{ffis2021}} 

\noindent The product Duality SecurePlus\textsuperscript{TM} Query enables financial crime and 
compliance teams to carry out collaborative investigations across groups, 
institutions and borders by keeping sensitive data encrypted throughout 
the process. 
This product allows cross-border collaboration between financial crime and compliance
teams. The product allows participant to query distributed data silos while keeping 
the sensitive data encrypted throughout the querying process. 
A case study between several banks in the UK and Canada demonstrated the capability 
of banks to match customers that includes external data sets, 
without disclosing their original query to the data owner. 
This allows different financial intelligence teams to securely query 
each other's data. 
The system also supports to reduce false positives, improve outcomes, 
increase efficiency, and most importantly effectively expose financial crime networks.\footnote{\href{https://dualitytech.com/use-cases/financial-services-industry/anti-money-laundering-aml/}{Leverage Secure Data Sharing to Fight Money Laundering}}

\paragraph*{DANIE Consortium~\cite{ffis2021}}

\noindent Project DANIE is a collaborative initiative involving multiple financial institutions from EU Member States, aimed at enhancing banking data quality through
the utilisation of privacy technologies. The project employs a combination of secure
hardware and software components, along with secure multiparty computation and 
homomorphic encryption. The primary objective of DANIE is to improve data quality by 
matching datasets across participating organisations to identify discrepancies and 
outliers, while simultaneously preserving the anonymity and security of the matched 
data through these technologies. In addition to its application in enhancing data reconciliation and subsequently improving the Know Your Customer (KYC) process, DANIE 
has the potential to be adapted for addressing the Suspicious Person List (SPL) 
cases discussed in \Cref{subsec:spl}. This can be achieved by encrypting an SPL
and uploading it to the DANIE platform to request confirmation of each 
individual's presence within the participating financial institutions.

\paragraph*{BNY Mellon and Inpher XOR~\cite{InpherTechreport}}

\noindent 
The Bank of New York Mellon Corporation uses machine-learning (ML) based
fraud detection algorithms to find hidden patterns in money laundering
networks.  
Together with the Inpher's XOR platform, BNY Mellon built a novel
collaborative fraud detection framework leveraging Secure Multiparty 
Computation. Inpher's technology enabled BNY Mellon to access  
sensitive data across borders or jurisdictional silos for training 
ML algorithms such as linear and logistic regression as well
as XGBoost. The solution achieved a high level of precision, computing 
models identically as if they were computed in clear text. 
BNY Mellon's project demonstrated that the number of false positives was
reduced by 20\% and the number of true positives was increased by 20\%. 
The solution thus reduced investigation workload and increased fraud 
detection efficiency.

\paragraph*{UK Tri-bank initiative~\cite{ffis2024}}

\noindent 
This is a proof-of-concept (POC) system developed by three UK banks to 
analyse historical data on small and medium sized corporate client datasets. 
The system used the FutureFlow Cross-Bank Financial Crime Analytics 
Platform to analyse transaction data between banks. In their POC
study, banks analysed approximately 200,000 accounts for money laundering
risk. In this system, the transaction data is encrypted to protect 
sensitive data before using to identify links between the transactions. 
This allowed the banks to identify potential money laundering networks 
based on the patterns in the transaction data.

\paragraph*{AUSTRAC's Fintel Alliance Alerting Project~\cite{chadderton2019}}

\noindent The FINTEL Alliance comprises representatives from both public 
and private partner organizations that collaborate to address complex or
emerging crimes affecting the community, necessitating a joint public-private
approach. 
The aim of this project is to develop different approaches which would allow
the identification of complex money transferring networks used by criminal 
gangs. 
The project used a federated platform with homomorphic encryption technologies
where entities could report or provide 
access to the agreed dataset via an API. The project was tested in a simulated
federated architecture using a sample of real data, covering more than 100 million 
accounts.

\paragraph*{AML Bridge Platform~\cite{fatf2022}}

\noindent AML Bridge is a secure platform provided by Salv. 
The platform uses an end-to end encryption technique which allows financial 
institutions to securely exchange of pseudonymised transaction data.
The platform was first tested as a POC on four of Estonia's largest banks 
and later expanded to all Estonian banks.
In their POC data owners can also limit what queries it allows
from its own data sources while the requesting party is not revealed 
to the data owner. In their querying protocol, the platform first 
encrypts the data at the source. Next data is aggregated with other parties' 
and stored in three trusted hosts using secure multiparty computation. 
Finally, within the requesting party's trusted environment, 
the platform aggregates the data results without revealing the 
original data source(s). 

\paragraph*{Transaction Monitoring Netherlands (TMNL)~\cite{van2024}}

\noindent This is an initiative consists of five Dutch banks including
ABN AMRO, ING, Rabobank, Triodos Bank and De Volksbank. The main aim of this 
initiative is to combine transaction data of business accounts from these
banks to gain a complete
view of money flows by criminal networks. The initiative uses a central trusted 
data store where each bank could upload their data. The data could then aggregate
by combining these transactions and the consortium (the group of banks) could 
monitor and report potentially suspicious patterns to the others. 
central data store where the consortium combines these transactions, monitors 
them, and reports potentially suspicious patterns. 
The central data store encrypts data at rest, which can only be viewed when
when suspicions of money laundering or terrorist financing arise
Sensitive information is 
encrypted and viewed only when suspicions of money laundering or terrorist 
financing arise. 

\paragraph*{DeepProtect~\cite{kanamori2022}} 

\noindent 
This is research project by National Institute of 
Information and Communications Technology (NICT) in Japan.
In this project, NICT collaborated with five banks in Japan 
including Chiba Bank, Ltd., MUFG Bank, Ltd., the Chugoku Bank, Ltd., 
Sumitomo Mitsui Trust Bank, Ltd., and the Iyo Bank, Ltd. 
The aim of this project was to build models for detecting criminals' bank
accounts and fraudulent transactions in customers accounts. 
DeepProtect employs a FL framework which enables each bank to run 
a local model using its own data while collaboratively train a global 
model with data from other banks. For privacy preservation, the implementation
used a homomorphic encryption scheme in a symmetric key environment. 

\paragraph*{Project Aurora~\cite{AuroraTechreport}} 

\noindent This is an initiative led by the Bank for International
Settlements (BIS). The aim of this project is to identify money
laundering between banks and cross-border financial institutions
by analysing the connected payment data. The project 
uses different privacy and security techniques with ML and network
analysis methods for identifying potential suspicious transactions. 
Phase 1 of the project demonstrated that the combinations of these technologies
provides more effective identification of money laundering networks compared to 
traditional siloed and rules-based approaches. In their experimental setting, 
consisting 29 financial institutions in six different European 
countries, the project detected potentially up to 3 times more money laundering involved 
complex schemes and reduced false positives by up to 80\%. 

\paragraph*{COSMIC~\cite{CosmicTechreport}} 

\noindent 
COSMIC is a secure digital platform co-developed by Monetary Authority of 
Singapore (MAS) and six major commercial banks in Singapore including DBS, OCBC, 
UOB, Citibank, HSBC and Standard Chartered.
Released in 2024, COSMIC allows financial institutions to securely share transaction 
data and information about suspicious accounts with one another. 
The initial aim of COSMIC is to provide a easier and efficient data sharing platform
for FIs to detect and thereby deter criminal activity. Currently, information 
sharing in COSMIC is voluntary. The platform currently supports three key 
financial crime risks, (1) misuse of legal persons, (2) misuse of trade 
finance for illicit purposes and (3) proliferation financing. COSMIC is owned 
and operated by the MAS.
\newline

\noindent The above shortlist of projects is quite compelling and is demonstrative of the increasing global interest in the adoption of maturing privacy technologies by major financial institutions, both public and private, to combat financial crimes.

\section{Discussion}\label{sec:discussion}
\paragraph*{Legislative Reform and Privacy Technologies}
Financial intelligence equips law enforcement and financial institutions with sophisticated analytical tools and data-driven methodologies to proactively detect and prevent financial crimes. 
While financial intelligence plays a crucial role in supporting Anti-Money Laundering (AML) and Counter-Terrorist Financing (CTF) regimes by delivering actionable intelligence and insights, 
it may present challenges in relation to data privacy regulations such as Australia's Privacy Act and European Union's General Data Protection Regulation (GDPR). 
These challenges arise particularly when there is a discrepancy between the initial purpose for which data was collected by financial institutions and its subsequent use in various financial intelligence applications. While public and private sectors, industry associations and regulatory bodies are actively working to harmonise AML/CTF and data privacy regulations, such changes often involve lengthy and uncertain political approval and implementation processes. 

From the practical perspective of being able to do something about complex financial crimes \emph{today}, we advocate for the integration of privacy-preserving technologies within financial intelligence tools, not just as a mitigation strategy while waiting for legal experts to navigate the complex AML/CTF reform landscape \cite{bugeja2018aml}, but also to inform those AML/CTF reform discussions. 
This approach offers several benefits. First, privacy-preserving technologies can be swiftly implemented, allowing financial institutions and financial intelligence units to address data privacy concerns while fulfilling AML/CTF obligations. Second, adopting privacy-preserving methods helps institutions build trust with customers and stakeholders by demonstrating a commitment to both data protection and effective financial crime prevention. Finally, privacy-preserving solutions offer a flexible framework that can adapt to evolving regulatory requirements, helping institutions stay compliant as laws and regulations are updated to address new financial technologies and emerging criminal threats.

\paragraph*{Trade-offs and Practical Considerations}
In practical financial intelligence applications, selecting the most appropriate 
privacy techniques is crucial for ensuring data security, regulatory compliance, and 
maintaining user trust at sustainable costs.
There are often trade-offs involved in such choices. 
With the increasing industry focus on responsible AI practices \cite{dignum2019responsible, kapoor2023reforms}, there is also an emerging need to understand the possible interactions between privacy requirements and related issues like explainability and fairness \cite{bagdasaryan2019differential, pujol2020fair, tran2021differentially}. 
Scholars and practitioners must look ahead to how AI itself may transform AML/CTF, potentially enabling new, automated, or hybrid methods especially when combined with emerging technologies such as decentralised autonomous organisations~\cite{ibalanky2025}. Strategic foresight and scenario-based methodologies may therefore be essential to anticipate and prepare for the future evolution of AI-enabled AML/CTF. To inform those broader issues, we offer here a few observations on how privacy-utility trade-offs are usually handled for the different privacy technologies. 

First of all, it is worth noting that the use of homomorphic encryption (HE) and secure multiparty computation (MPC) in financial-intelligence algorithms like private set intersection, graph pattern matching, and statistical model learning do not usually result in any change to the output of those algorithms. This is because HE and MPC allow an algorithm to be evaluated in an encrypted space but they do not otherwise change the nature of a given algorithm.
The primary consideration when using HE and MPC is thus whether the specific algorithm and parameters chosen provide, at reasonable organisational and monetary costs, sufficient provable security against adversaries with access to potentially large computational resources and arbitrary background and side information.

In contrast, the use of differential privacy (DP) and federated learning (FL) to privatise a financial-intelligence algorithm can result in actual changes to the output of the algorithm.
There are rich and well-established frameworks \cite{dwork2014algorithmic, kifer2014pufferfish, nissim2021privacy} to study trade-offs between privacy and utility when using DP, and these frameworks work well in conjunction with standard techniques for model evaluation and selection \cite{raschka2018model} in both federated learning and more typical setups. Addressing these questions will be central to future research in this field.

\section{Conclusion}

We hope this paper enhances public understanding of privacy-preserving financial intelligence through the various AML/CTF use cases we explore. The use cases highlight the evolving nature of financial crimes and the necessity for ongoing innovation in privacy technologies to address emerging threats effectively. Having said that, for financial institutions and regulatory bodies, adopting privacy technologies involves careful consideration of both technological and operational aspects. Effective implementation strategies must balance the trade-offs between data privacy and the quality and efficiency of analysis.
While these technologies address immediate challenges, they are not a substitute for comprehensive regulatory reforms. 
Further and continuous refinement of regulations is essential to align financial crime prevention efforts with data privacy goals.

%

\paragraph*{Acknowledgements}
We thank the many former colleagues in AUSTRAC and financial crime units in major Australian financial institutions who spent many hours teaching us the intricacies of the financial intelligence business. We are grateful to many of them for continuing to play an important role to keep our financial system and our country safe.

\bibliographystyle{elsarticle-num} 
\bibliography{cas-refs}

@article{tran2021differentially,
  title={Differentially private empirical risk minimization under the fairness lens},
  author={Tran, Cuong and Dinh, My and Fioretto, Ferdinando},
  journal={Advances in Neural Information Processing Systems},
  volume={34},
  pages={27555--27565},
  year={2021}
}

@article{bagdasaryan2019differential,
  title={Differential privacy has disparate impact on model accuracy},
  author={Bagdasaryan, Eugene and Poursaeed, Omid and Shmatikov, Vitaly},
  journal={Advances in Neural Information Processing Systems},
  volume={32},
  year={2019}
}

@inproceedings{pujol2020fair,
  title={Fair decision making using privacy-protected data},
  author={Pujol, David and McKenna, Ryan and Kuppam, Satya and Hay, Michael and Machanavajjhala, Ashwin and Miklau, Gerome},
  booktitle={Proceedings of the Conference on Fairness, Accountability, and Transparency},
  pages={189--199},
  year={2020}
}

@article{raschka2018model,
  title={Model evaluation, model selection, and algorithm selection in machine learning},
  author={Raschka, Sebastian},
  journal={arXiv:1811.12808},
  year={2018}
}

@book{dignum2019responsible,
  title={Responsible artificial intelligence: how to develop and use AI in a responsible way},
  author={Dignum, Virginia},
  year={2019},
  publisher={Springer}
}

@article{kapoor2023reforms,
  title={Reforms: Reporting standards for machine learning based science},
  author={Kapoor, Sayash and Cantrell, Emily and Peng, Kenny and Pham, Thanh Hien and Bail, Christopher A and Gundersen, Odd Erik and Hofman, Jake M and Hullman, Jessica and Lones, Michael A and Malik, Momin M},
  journal={arXiv:2308.07832},
  year={2023}
}

@article{yousuf2021systematic,
  title={A systematic review of {C}rypt{DB}: Implementation, challenges, and future opportunities},
  author={Yousuf, Hana and Salloum, S and Aburayya, A and Al-Emran, M and Shaalan, K},
  journal={J. Manag. Inf. Decis. Sci},
  volume={24},
  number={1},
  pages={1--16},
  year={2021}
}

@inproceedings{popa2011cryptdb,
  title={Crypt{DB}: Protecting confidentiality with encrypted query processing},
  author={Popa, Raluca Ada and Redfield, Catherine MS and Zeldovich, Nickolai and Balakrishnan, Hari},
  booktitle={Proceedings of the 23rd ACM Symposium on Operating Systems Principles},
  pages={85--100},
  year={2011}
}

@article{mothukuri2021survey,
  title={A survey on security and privacy of federated learning},
  author={Mothukuri, Viraaji and Parizi, Reza M and Pouriyeh, Seyedamin and Huang, Yan and Dehghantanha, Ali and Srivastava, Gautam},
  journal={Future Generation Computer Systems},
  volume={115},
  pages={619--640},
  year={2021},
  publisher={Elsevier}
}

@article{burkhardt2021optimal,
  title={Optimal algebraic Breadth-First Search for sparse graphs},
  author={Burkhardt, Paul},
  journal={ACM Transactions on Knowledge Discovery from Data},
  volume={15},
  number={5},
  pages={1--19},
  year={2021},
  publisher={ACM}
}

@article{kifer2014pufferfish,
  title={Pufferfish: A framework for mathematical privacy definitions},
  author={Kifer, Daniel and Machanavajjhala, Ashwin},
  journal={ACM Transactions on Database Systems (TODS)},
  volume={39},
  number={1},
  pages={1--36},
  year={2014},
  publisher={ACM}
}

@techreport{alagic2022status,
    author = {Alagic, Gorjan and others},
    title = {Status report on the third round of the {NIST} post-quantum cryptography standardization process},
    institution = {NIST},
    year = 2022
}

@book{gilmore04,
    author = "William C. Gilmore",
    title = "Dirty Money: The Evolution of International Measures to Counter Money Laundering and the Financing of Terrorism",
    publisher = "Council of Europe Publishing",
    edition = "3rd",
    year = 2004,
    address = {Strasbourg}
}

@book{reuter04,
    author = "Peter Reuter and Edwin M. Truman",
    title = "Chasing Dirty Money: The Fight Against Money Laundering",
    publisher = "Institute for International Economics",
    year = 2004,
address={Washington}
}

@book{leuprecht23,
    editor = "Christian Leuprecht and Jamie Ferrill",
    title = "Dirty Money: Financial Crime in Canada",
    publisher = "McGill-Queen's University Press",
    year = 2023,
    address = {Kingston, ON}
}

@book{madinger2011money,
  title={Money laundering: A guide for criminal investigators},
  author={Madinger, John},
  year={2011},
  publisher={CRC Press},
  address={Boca Raton, Florida}
}

@inproceedings{nissim2021privacy,
  title={Privacy: From database reconstruction to legal theorems},
  author={Nissim, Kobbi},
  booktitle={Principles of Database Systems},
  pages={33--41},
  year={2021}
}

@article{dwork2014algorithmic,
  title={The Algorithmic Foundations of Differential Privacy},
  author={Dwork, Cynthia and Roth, Aaron},
  journal={Foundations and Trends{\textregistered} in Theoretical Computer Science},
  volume={9},
  number={3--4},
  pages={211--407},
  year={2014},
  publisher={Now Publishers}
}

@book{lynch22,
  author = {Nathan Lynch},
  title = {The Lucky Laundry},
  publisher = {Harper Collins},
  year = {2022},
  address = {Syndey, NSW}
}

@article{kearns2016private,
  title={Private algorithms for the protected in social network search},
  author={Kearns, Michael and Roth, Aaron and Wu, Zhiwei Steven and Yaroslavtsev, Grigory},
  journal={Proceedings of the National Academy of Sciences},
  volume={113},
  number={4},
  pages={913--918},
  year={2016},
  publisher={National Acad Sciences}
}

@inproceedings{dwork2007price,
  title={The price of privacy and the limits of {LP} decoding},
  author={Dwork, Cynthia and McSherry, Frank and Talwar, Kunal},
  booktitle={ACM symposium on Theory of Computing},
  pages={85--94},
  year={2007}
}

@inproceedings{dinur2003revealing,
  title={Revealing information while preserving privacy},
  author={Dinur, Irit and Nissim, Kobbi},
  booktitle={Proceedings of the ACM Symposium on Principles of Database Systems},
  pages={202--210},
  year={2003}
}

@article{brand2023nonlinear,
  title={Nonlinear computations on {F}in{T}racer tags},
  author={Brand, Michael and Churchill, Tania and Friedrich, Carsten},
  journal={Cryptology ePrint Archive},
  year={2023}
}

@inproceedings{purcell2023split,
  title={Split, count, and share: a differentially private set intersection cardinality estimation protocol},
  author={Purcell, Michael and Li, Yang and Ng, Kee Siong},
  booktitle={Uncertainty in Artificial Intelligence},
  pages={1684--1694},
  year={2023},
  publisher={PMLR}
}

@article{mouzakiti2020cooperation,
  title={Cooperation Between Financial Intelligence Units in the {EU}: Stuck in the Middle Between the {GDPR} and the Police Data Protection Directive},
  author={Mouzakiti, Foivi},
  journal={New Journal of European Criminal Law},
  year={2020}
}

@article{balboni2013privacy,
  title={Privacy by design and anonymisation techniques in action: {C}ase study of {M}a3tch technology},
  author={Balboni, Paolo and Macenaite, Milda},
  journal={Computer Law \& Security Review},
  volume={29},
  number={4},
  pages={330--340},
  year={2013},
  publisher={Elsevier}
}

@book{christenRS20,
  author    = {Peter Christen and
               Thilina Ranbaduge and
               Rainer Schnell},
  title     = {Linking Sensitive Data: Methods and Techniques for Practical Privacy-Preserving
               Information Sharing},
  publisher = {Springer},
  year      = {2020}
}

@inproceedings{zhang18,
  title={Scalable entity resolution using probabilistic signatures on parallel databases},
  author={Zhang, Yuhang and Ng, Kee Siong and Churchill, Tania and Christen, Peter},
  booktitle={Proceedings of the 27th ACM International Conference on Information and Knowledge Management},
  pages={2213--2221},
  year={2018}
}

@inproceedings{ zhang17,
  author    = {Yuhang Zhang and
               Tania Churchill and
               Kee Siong Ng},
  title     = {Exploiting Redundancy, Recurrency and Parallelism: How to Link Millions
               of Addresses with Ten Lines of Code in Ten Minutes},
  booktitle = {Australasian Conference on Data Mining},
  pages     = {107--122},
  publisher = {Springer},
  year      = {2017}
}

@article{ savageWCZY16,
  title={Detection of money laundering groups using supervised learning in networks},
  author={Savage, David and Wang, Qingmai and Chou, Pauline and Zhang, Xiuzhen and Yu, Xinghuo},
  journal={arXiv:1608.00708},
  year={2016}
}

@inproceedings{savageWZCY17,
  author    = {David Savage and
               Qingmai Wang and
               Xiuzhen Zhang and
               Pauline Lienhua Chou and
               Xinghuo Yu},
  title     = {Detection of Money Laundering Groups: Supervised Learning on Small
               Networks},
  booktitle = {{AI} and {OR} for Social Good, Papers from the {AAAI} Workshop},
  year      = {2017}
}

@article{ savage-tpds-17,
  author    = {David Savage and
               Xiuzhen Zhang and
               Pauline Chou and
               Xinghuo Yu and
               Qingmai Wang},
  title     = {Distributed Mining of Contrast Patterns},
  journal   = {{IEEE} Trans. Parallel Distributed Syst.},
  volume    = {28},
  number    = {7},
  pages     = {1881--1890},
  year      = {2017}
}

@article{ leuprecht18,
   author = "Christian Leuprecht and Arthur Crockfield and Pam Simpson and Maseeh Haseeb",
   title = "Tracking Transnational Terrorist Resourcing Nodes and Networks",
   journal = "Florida State University Law Review",
   year = 2019
}

@article{li21,
  title={Private graph data release: A survey},
  author={Li, Yang and Purcell, Michael and Rakotoarivelo, Thierry and Smith, David and Ranbaduge, Thilina and Ng, Kee Siong},
  journal={ACM Computing Surveys},
  volume={55},
  number={11},
  pages={1--39},
  year={2023},
  publisher={ACM}
}

@inproceedings{byrd2020,
  title={Differentially private secure multi-party computation for federated learning in financial applications},
  author={Byrd, David and Polychroniadou, Antigoni},
  booktitle={Proceedings of the First ACM International Conference on AI in Finance},
  pages={1--9},
  year={2020}
}

@inproceedings{sangers2019secure,
  title={Secure multiparty PageRank algorithm for collaborative fraud detection},
  author={Sangers, Alex and van Heesch, Maran and Attema, Thomas and Veugen, Thijs and Wiggerman, Mark and Veldsink, Jan and Bloemen, Oscar and Worm, Dani{\"e}l},
  booktitle={Financial Cryptography and Data Security: 23rd International Conference},
  pages={605--623},
  year={2019},
  publisher={Springer}
}

@article{brand2023fintracer,
  title={Fin{T}racer: A privacy-preserving mechanism for tracing electronic money},
  author={Brand, Michael and Ivey-Law, Hamish and Churchill, Tania},
  journal={Cryptology ePrint Archive},
  year={2023}
}

@inproceedings{gentry2009fully,
  title={Fully homomorphic encryption using ideal lattices},
  author={Gentry, Craig},
  booktitle={Proceedings of the 41st Annual ACM Symposium on Theory of Computing},
  pages={169--178},
  year={2009}
}

@article{chi2017,
  title={Hashing techniques: A survey and taxonomy},
  author={Chi, Lianhua and Zhu, Xingquan},
  journal={ACM Computing Surveys},
  volume={50},
  number={1},
  pages={1--36},
  year={2017},
  publisher={ACM}
}

@article{aledhari2020,
  title={Federated learning: A survey on enabling technologies, protocols, and applications},
  author={Aledhari, Mohammed and Razzak, Rehma and Parizi, Reza M and Saeed, Fahad},
  journal={IEEE Access},
  volume={8},
  pages={140699--140725},
  year={2020},
  publisher={IEEE}
}

@article{nguyen2021,
  title={Federated learning for internet of things: A comprehensive survey},
  author={Nguyen, Dinh C and Ding, Ming and Pathirana, Pubudu N and Seneviratne, Aruna and Li, Jun and Poor, H Vincent},
  journal={IEEE Communications Surveys \& Tutorials},
  volume={23},
  number={3},
  pages={1622--1658},
  year={2021},
  publisher={IEEE}
}

@misc{zihao2022,
  author = {Zhao, Zihao and Luo, Mengen and Ding, Wenbo},
  title = {Deep Leakage from Model in Federated Learning},
  publisher = {arXiv},
  year = {2022},
}

@inproceedings{rosulek2021compact,
  title={Compact and malicious private set intersection for small sets},
  author={Rosulek, Mike and Trieu, Ni},
  booktitle={Proceedings of the ACM SIGSAC Conference on Computer and Communications Security},
  pages={1166--1181},
  year={2021}
}

@inproceedings{chase2020private,
  title={{Private set intersection in the internet setting from lightweight oblivious PRF}},
  author={Chase, Melissa and Miao, Peihan},
  booktitle={Annual International Cryptology Conference},
  pages={34--63},
  year={2020},
  publisher={Springer}
}

@inproceedings{chen2017fast,
  title={Fast private set intersection from homomorphic encryption},
  author={Chen, Hao and Laine, Kim and Rindal, Peter},
  booktitle={Proceedings of the ACM SIGSAC Conference on Computer and Communications Security},
  pages={1243--1255},
  year={2017}
}

@inproceedings{freedman2004efficient,
  title={Efficient private matching and set intersection},
  author={Freedman, Michael J and Nissim, Kobbi and Pinkas, Benny},
  booktitle={International Conference on the Theory and Applications of Cryptographic Techniques},
  pages={1--19},
  year={2004},
  publisher={Springer}
}

@inproceedings{molloy2017graph,
  title={Graph analytics for real-time scoring of cross-channel transactional fraud},
  author={Molloy, Ian and Chari, Suresh and Finkler, Ulrich and Wiggerman, Mark and Jonker, Coen and Habeck, Ted and Park, Youngja and Jordens, Frank and van Schaik, Ron},
  booktitle={20th International Conference on Financial Cryptography and Data Security},
  pages={22--40},
  year={2017},
  publisher={Springer}
}

@inproceedings{dhiman2023,
  title={Homomorphic encryption based federated learning for financial data security},
  author={Dhiman, Shalini and Nayak, Sumitra and Mahato, Ganesh Kumar and Ram, Anil and Chakraborty, Swarnendu Kumar},
  booktitle={4th International Conference on Computing and Communication Systems},
  pages={1--6},
  year={2023},
  publisher={IEEE}
}

@incollection{suzumura2022,
  title={Federated learning for collaborative financial crimes detection},
  author={Suzumura, Toyotaro and Zhou, Yi and Kawahara, Ryo and Baracaldo, Nathalie and Ludwig, Heiko},
  booktitle={Federated learning: A comprehensive overview of methods and applications},
  pages={455--466},
  year={2022},
  publisher={Springer}
}

@article{elgamal1985public,
  title={A public key cryptosystem and a signature scheme based on discrete logarithms},
  author={ElGamal, Taher},
  journal={IEEE Transactions on Information Theory},
  volume={31},
  number={4},
  pages={469--472},
  year={1985},
  publisher={IEEE}
}

@inproceedings{bernstein2006curve25519,
  title={Curve25519: new {D}iffie-{H}ellman speed records},
  author={Bernstein, Daniel J},
  booktitle={9th International Conference on Theory and Practice in Public-Key Cryptography},
  pages={207--228},
  year={2006},
  publisher={Springer}
}

@misc{ndis2020,
  author = {AUSTRAC},
  title = {National Disability Insurance Scheme Fraud Prevention Financial Crime Guide},
  howpublished = {\url{https://www.austrac.gov.au/sites/default/files/2020-12/NDIS%20Fraud%20Prevention%20Financial%20Crime%20Guide.pdf}},
  year = 2020
}

@article{hardy2017,
  title={Private federated learning on vertically partitioned data via entity resolution and additively homomorphic encryption},
  author={Hardy, Stephen and Henecka, Wilko and Ivey-Law, Hamish and Nock, Richard and Patrini, Giorgio and Smith, Guillaume and Thorne, Brian},
  journal={arXiv:1711.10677},
  year={2017}
}

@article{bugeja2018aml,
  title={{AML/CFT} and data privacy regulation: Achieving a peaceful co-existence},
  author={Bugeja, Diane},
  journal={Journal of Financial Compliance},
  volume={2},
  number={2},
  pages={132--141},
  year={2018},
  publisher={Henry Stewart Publications}
}

@article{van2024privacy,
  title={Privacy-preserving Anti-Money Laundering using Secure Multi-Party Computation},
  author={van Egmond, Marie Beth and Dunning, Vincent and van den Berg, Stefan and Rooijakkers, Thomas and Sangers, Alex and Poppe, Ton and Veldsink, Jan},
  journal={Cryptology ePrint Archive},
  year={2024}
}

@article{kanamori2022,
  title={Privacy-preserving federated learning for detecting fraudulent financial transactions in {J}apanese banks},
  author={Kanamori, Sachiko and Abe, Taeko and Ito, Takuma and Emura, Keita and Wang, Lihua and Yamamoto, Shuntaro and Le, Trieu Phong and Abe, Kaien and Kim, Sangwook and Nojima, Ryo},
  journal={Journal of Information Processing},
  volume={30},
  pages={789--795},
  year={2022},
  publisher={Information Processing Society of Japan}
}

@techreport{AuroraTechreport,
  title       = {Project {A}urora: The power of data, technology and collaboration to combat money laundering},
  institution      = {Bank for International Settlements},
  author = {BIS},
  year        = {2023}
}

@techreport{InpherTechreport,
  title       = {New Approaches to Tackling Financial Crimes: Secure Data Collaboration With {XOR}},
  author      = {Inpher}, 
  institution = {BNY Mellon},
  year        = {2021}
}

@article{ffis2021,
  title={Innovation and discussion paper: Case studies of the use of privacy preserving analysis to tackle financial crime},
  author={Nick Maxwell},
  journal={Future of Financial Intelligence Sharing (FFIS) research programme},
  year={2021}
}

@article{ffis2024,
  title={Paper 1: The case for national policy-makers to unleash the potential of payments infrastructure to identify economic crime risk},
  author={Nick Maxwell},
  journal={Future of Financial Intelligence Sharing (FFIS) research programme},
  year={2024},
  url = {https://www.future-fis.com/uploads/3/7/9/4/3794525/ffis_-_payments_policy_discussion_paper_1_-_unleashing_the_potential_of_payments_infrastructure_to_identify_economic_crime_risk.pdf} 
}

@article{fatf2022,
  title={Partnering in the Fight Against Financial Crime: Data Protection, Technology and Private
Sector Information Sharing},
  author={FATF},
  journal={The Financial Action Task Force (FATF)},
  address={France},
  year={2022},
  url={https://www.fatf-gafi.org/content/dam/fatf-gafi/guidance/Partnering-int-the-fight-against-financial-crime.pdf}
}

@article{van2024,
  title={Privacy-preserving Anti-Money Laundering using Secure Multi-Party Computation},
  author={van Egmond, Marie Beth and Dunning, Vincent and van den Berg, Stefan and Rooijakkers, Thomas and Sangers, Alex and Poppe, Ton and Veldsink, Jan},
  journal={Cryptology ePrint Archive},
  year={2024}
}

@techreport{chadderton2019,
  title={Public-private partnerships to disrupt financial crime: An exploratory study of {A}ustralia’s {F}intel {A}lliance},
  author={Chadderton, Paula and Norton, Simon},
  year={2019},
  institution={SWIFT Institute Working Paper}
}

@article{scott2024,
  title={Anti-Money Laundering/Counter-Terrorism Financing Tranche 2 reform in {A}ustralia - {A}n opportunity for intelligence to lead the way},
  author={Scott, Benjamin and Webster, Mark},
  journal={International Journal of Contemporary Intelligence Issues},
  volume={1},
  number={1},
  pages={3--19},
  year={2024},
  publisher={Australian Institute of Professional Intelligence Officers Deakin West, ACT}
}

@techreport{CosmicTechreport,
  title       = {Industry Perspectives on Best Practices - {L}everaging on Data Analytics and Machine Learning Methods for {AML/CFT} },
  author      = {{ABS}},
  institution = {The Association of Banks of Singapore},
  year        = {2024}
}

@article{ibalanky2025,
  title={Applying AI to Canada's Financial Intelligence System: Promises and Perils in Combatting Money Laundering and Terrorism Financing},
  author={Ibalanky, Corinne and Wilner, Alex},
  journal={International Journal},
  volume={80},
  number={2},
  pages={147--165},
  year={2025},
  publisher={SAGE Publications Sage UK: London, England}
}

\end{document}